\documentclass[twocolumn]{aastex63}
\usepackage{amsmath}
\usepackage{amsfonts}
\usepackage{amssymb}
\usepackage{bm}
\usepackage{color}
\usepackage{graphicx}
\usepackage{natbib}
\usepackage[utf8]{inputenc}

\shorttitle{Bright-Point Shape Changes}
\shortauthors{Van Kooten \& Cranmer}

\accepted{January 7, 2024}
\submitjournal{\apj}

\newcommand{\muram}{MURaM}
\newcommand{\Al}{Alfv\'en}
\newcommand{\BP}{Bright Point}
\newcommand{\Bp}{Bright point}
\newcommand{\bp}{bright point}
\newcommand{\bhp}{bright-point}

\newcommand{\BHP}{Bright-Point}
\renewcommand{\max}{\ensuremath{_\text{max}}}
\renewcommand{\min}{\ensuremath{_\text{min}}}
\renewcommand{\vec}[1]{\ensuremath{\boldsymbol{\mathbf{#1}}}}

\newcommand{\sd}{\sigma}

\begin{document}

\title{Using Bright-Point Shapes to Constrain Wave-Heating of the Solar Corona: Predictions for DKIST}

\author[0000-0002-4472-8517]{Samuel J. Van Kooten}
\affil{Southwest Research Institute, Boulder, Colorado, USA}
\affil{Department of Astrophysical and Planetary Sciences, University of Colorado, Boulder, Colorado, USA}

\author[0000-0002-3699-3134]{Steven R. Cranmer}
\affil{Department of Astrophysical and Planetary Sciences, University of Colorado, Boulder, Colorado, USA}

\correspondingauthor{Samuel Van Kooten}
\email{samuel.vankooten@swri.org}

\begin{abstract}
	Magnetic bright points on the solar photosphere mark the footpoints of kilogauss magnetic flux tubes extending toward the corona.
	Convective buffeting of these tubes is believed to excite magnetohydrodynamic waves, which can propagate to the corona and there deposit heat.
	Measuring wave excitation via bright-point motion can thus constrain coronal and heliospheric models, and this has been done extensively with centroid tracking, which can estimate kink-mode wave excitation.
	DKIST is the first telescope to provide well-resolved observations of bright points, allowing shape and size measurements to probe the excitation of other wave modes that have been difficult, if not impossible, to study to date.
	In this work, we demonstrate a method of automatic \bhp\ tracking that robustly identifies the shapes of \bp s, and we develop a technique for interpreting measured \bhp\ shape changes as the driving of a range of thin-tube wave modes.
	We demonstrate these techniques on a \muram\ simulation of DKIST-like resolution.
	These initial results suggest that modes other than the long-studied kink mode could increase the total available energy budget for wave-heating by 50\%.
	Pending observational verification as well as modeling of the propagation and dissipation of these additional wave modes, this could represent a significant increase in the potency of wave--turbulence heating models.
\end{abstract}


\section{Introduction}
\label{sec:intro}

Photospheric bright points are intensity enhancements of order 100~km in size and with lifetimes of order 5~min \citep{Berger1995,Goode2010,Liu2018}, seen in the intergranular lanes of the photosphere, with greater densities along the supergranular network and in plage regions.
\Bp s usually correspond with kilogauss concentrations of vertical magnetic flux \citep{Utz2013}, sometimes called \textit{flux elements}, which are believed to be the bases of flux tubes that may rise to the corona, making them the source of much of the open magnetic flux in coronal holes \citep{Hofmeister2019}.
The rise of these flux tubes to the low chromosphere has been observed as brightenings in the chromospheric line Ca II H coinciding with photospheric \bp s \citep{Xiong2017,Liu2018a,Narang2019}.
As they rise, these flux tubes expand rapidly \citep{Kuckein2019} due to the
decreasing plasma pressure \citep{Spruit1976}.
These magnetic concentrations are responsible for the appearance of \bp s, as
the magnetic pressure reduces the plasma density, allowing emission from
slightly deeper, hotter layers.
This produces continuum enhancement and also much greater enhancement in the
G-band and other spectral lines \citep[e.g.][]{Uitenbroek2006}.

\Bp s are seen to be in nearly-continuous motion, buffeted by the
constantly-evolving granular pattern
\citep{VanBallegooijen1998,Nisenson2003,Utz2010,Chitta2012}, and this motion is
believed to excite flux tube waves which propagate up the tube to the corona,
where they can deposit heat through turbulent dissipation
\citep{Cranmer2005,Soler2019,Cranmer2019}.
Evidence of oscillations in the chromosphere has been observed above magnetic elements, e.g. by \citealp{Jafarzadeh2017a,Stangalini2017}, supporting this model.
This mechanism of wave-driving is proposed as one potential explanation for the driving of high temperatures in the solar corona.
A number of studies have used \bhp\ motions as a proxy for the driving of transverse, kink-mode waves in observations \citep[e.g.][]{Nisenson2003,Yang2014} and simulations \citep[e.g.][]{Keys2011} and have modeled the resulting waves \citep[e.g.][]{Cranmer2005,DeMoortel2022}, but each has been limited by the barely-resolved nature of these small \bp s.
Therefore, while kink-mode waves, associated with the overall centroid motion of the \bp, have been studied extensively, other wave modes associated with the shape of the flux-tube cross section are yet under-explored.
While efforts have been made to probe wave modes associated with \bhp\ area changes \citep[e.g.][]{Gao2021}, observations have been very limited in the wave modes than can be studied.

The new Daniel K. Inouye Solar Telescope (DKIST; \citealp{Rimmele2020,Rast2021}) is now changing this situation.
Its Visible Broadband Imager (VBI; \citealp{Woger2021}) achieves the highest spatial resolution ever for solar observations: a diffraction-limited, G-band resolution of 15~km on the photosphere, which can reveal the shapes and sub-structure of the $\sim$~100~km \bp s.
Paired with the instrument's high cadence (a few seconds), it is also possible to probe \bhp\ motions at very high frequencies---the frequencies most relevant to wave--turbulence heating models \citep[e.g.][]{Pelouze2023}.
Additionally, DKIST's spectropolarimetric instruments, including the Diffraction-Limited Near-Infrared Spectro-Polarimeter (DL-NIRSP; \citealp{Jaeggli2022}) and the Visible SpectroPolarimeter (ViSP; \citealp{deWijn2022}) provide unprecedented resolution and sensitivity, allowing detailed, small-scale probes into the magnetic structure of \bp s.

We propose a new technique for connecting observations of \bhp\ shape evolution to the driving of thin-tube waves in the overlying flux tubes.
Our overarching goals in the present work are to produce an initial estimate of the energy flux which could not be measured before DKIST, in order to determine whether further study is warranted (we find that it is), and to propose ideas which may inspire further development.
We demonstrate this technique, as well as a \bhp\ identification algorithm that robustly identifies \bhp\ shapes, on a \muram\ simulation of DKIST-like resolution, and we use this demonstration to motivate future work.

\section{Bright Point Identification}

\subsection{\muram\ Simulation}
\label{sec:2s-muram}
To demonstrate and analyze the techniques presented in the following sections, we use a \muram\ simulation of DKIST-like resolution.
\citet{Rempel2014} used an extensively-modified version of \muram\
\citep{Vogler2005,Rempel2009}, a radiative magnetohydrodynamic (MHD) simulation
code.
We use the outputs of a \muram\ run similar to the O16b run of \citet{Rempel2014} but with an expanded vertical domain above the photosphere and with radiative transfer computed in four opacity bins.
This run was also analyzed by \citet{Agrawal2018} and the Appendix of \citet{VanKooten2017}.
The simulation domain is $6\times6\times4$~Mm$^3$ at a 16~km grid spacing, with the photosphere approximately 1.7~Mm below the top of the domain.
The snapshot cadence is two seconds over about one hour of simulated time.
The simulation includes convective flows at well-resolved granular scales, with \bp s arising naturally from a magnetic field produced by a small-scale dynamo.
The upward-directed, white-light intensity is computed through radiative transfer with four opacity bins and produces an analog of observational images which we use in our analysis.

\begin{figure*}[t]
	\centering
	\includegraphics{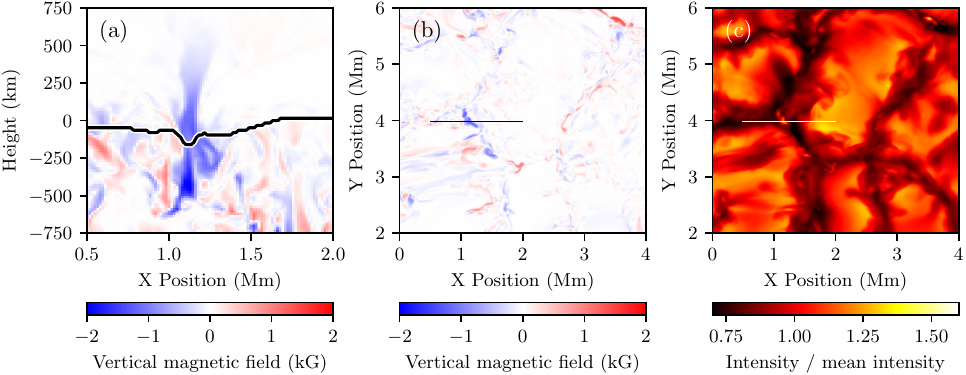}
	\caption{A sample \bp\ seen in the \muram\ simulation. Panel (a) shows the vertical magnetic field strength in a vertical slice through a \bp, with the $\tau=1$ surface marked by the black line. A strong concentration of vertical flux is seen, coinciding with a lowering of the optical surface by $\sim100$~km. The concentration drops rapidly with height as the flux tube expands. Panels (b) and (c) show the vertical field strength at the average $\tau=1$ level and the upward, white-light intensity, both for a subsection of the full domain. The granular pattern is seen in the intensity image, and vertical flux concentrations are confined to the downflow lanes. Horizontal lines in these plots indicate the location of the vertical slice in panel (a).}
	\label{fig:muram-sample}
\end{figure*}

We show sample data from the simulation in Figure~\ref{fig:muram-sample}.
It can be seen that small concentrations of vertical flux abound in the granular downflow lanes, and the strongest ones are associated with intensity enhancements.
The flux tube associated with a \bp\ is also shown.
The strength and verticality of the flux tube is greatest near the optical surface.
Deeper down, convective forces bend and twist the flux tube out of the plane of the plot.
Above the photosphere, the flux tube expands rapidly as the plasma pressure drops, reducing the field strength.

\subsection{\BHP\ Identification}
\label{sec:new-tracking-algo}

To identify \bp s, we employ a version of the tracking code of \citet{VanKooten2017}, modified by \citet{VanKooten2021a} to achieve robust identification of the edges of \bp s (rather than focusing only on centroid location). It consists of the following steps. (Values for all parameters are given in Table \ref{table:tracking_params}.)

\begin{table*}[t]
	\centering
	\begin{tabular}{lcc}
		Quantity & Value in previous work & Value in current work \\
		\hline
		$n_\sigma$ & 3 & 7.5 \\
		$n_\text{exp}$ & 3 & 9 \\
		$f_\text{contour}$ & --- & 0.65 \\
		$\Delta B\max$ & --- & 1.2 \\
		$f_\text{fp}$ & 0.2 & 0.2 \\
		$d_\text{prox}$ & 4~px & 2~px \\
		$A\min$ & 4~px$^2$ & 8~px$^2$ \\
		$A\max$ & 110~px$^2$ & 200~px$^2$ \\
		$d\max$ & 20~px & 30~px \\
		$t\min$ & 5 frames & 5 frames \\
		$\Delta s_\text{max,\%}$ & --- & 50\% \\
		$\Delta s_\text{max,px}$ & --- & 10~px \\
		Diagonal connections? & Yes & No \\
		
	\end{tabular}
	\caption{Tracking parameters used by \citet{VanKooten2017} and in the present work. The meaning of each quantity is explained in the text. Dashes indicate quantities not used in the previous work.}
	\label{table:tracking_params}
\end{table*}

\begin{enumerate}
\item We calculate the intensity of each pixel relative to the mean of its eight neighboring pixels (a discrete analog of the Laplacian), and we identify those pixels for which this value lies $n_\sigma$ standard deviations above the mean of this value (calculated across that frame).
This provides a collection of ``seed'' pixels which are significantly brighter than their immediate surroundings.
This is ideal for \bp s, which stand out strongly against the surrounding dark lanes but which are not necessarily the brightest features in the image.

\item Each contiguous set of seed pixels is expanded through $n_\textrm{exp}$ iterations to include eligible neighboring pixels.
For each set, we identify the maximum brightness $B\max$ of the constituent pixels as well as the minimum brightness $B\min$ of the pixels in a surrounding box (the minimal rectangle containing the seed pixels, widened in all directions by $n_\text{exp}$ pixels).
In each round of expansion, pixels are eligible to be added if their brightness exceeds $B\min + f_\text{contour} (B\max - B\min)$.
Effectively, expansion is constrained to be within a contour drawn at that threshold value, which is locally adapted to each \bp 's own contrast with its surroundings.
We also define a parameter $\Delta B\max$: when $B\max-B\min > \Delta B\max$, the contour is instead set at $B\min + f_\text{contour} \Delta B\max$.
This allows $f_\text{contour}$ to be kept at an appropriate level for dimmer \bp s without constricting the shape of the brightest \bp s by placing their contours too high.
A fixed number $n_\text{exp}$ of expansion rounds are conducted, and the neighboring pixels added to the feature in each round are added to the set of seed pixels for the next round of expansion.
The number of rounds of expansion is tuned to the typical pixel and \bhp\ sizes.

\item Following expansion, a ``false-positive'' rejection step eliminates features that would expand significantly in an additional expansion round.
These are typically features which are actually especially-bright portions of a granule, or \bp s without sufficiently-clear separation from surrounding granular pixels.
This is as opposed to true and clearly-identifiable \bp s, which are typically small and compact.
We calculate for each feature the fraction of its immediately surrounding pixels which would be added to the feature if another round of expansion were conducted.
If that fraction falls above a set threshold $f_\text{fp}$, the feature is rejected.

\item Features are also rejected if they touch the edge of the frame, if they are very large or very small (having an area $A$ $<A\min$ or $>A\max$, or a minimal bounding rectangle with a diagonal length exceeding $d\max$), or if they are extremely close to another \bp\ (within a distance $d_\text{prox}$) and thus possibly part of a larger, complex feature that was mis-identified as multiple fragments.

\item Remaining features are connected from frame to frame based on mutual overlap.
Merging and splitting events (i.e., a feature in one frame overlapping multiple features in the preceding or following frame) are treated as the end of the feature(s) going into the event and the beginning of the new feature(s) arising from the event.
If, between two frames, a feature's size changes by more than $\Delta s_\text{max,px}$ pixels and that change in area, measured relative to the smaller of the two sizes, is greater than $\Delta s_\text{max,\%}$, the feature is not connected across those two frames.
That is, the first feature is ended in the earlier frame, and the second, very differently-sized feature is regarded as beginning in the second frame.
This avoids very large impulsive velocities inferred when a sudden, large change in \bhp\ area occurs, which is usually due to a large number of pixels being very close to a relevant threshold rather than a large change in the underlying intensity distribution.
Manual inspection of a sample of such events supports this characterization.
A final lifetime criterion eliminates features that are tracked for fewer than $t\min$ seconds.

\end{enumerate}

In the present work, we tune many of the thresholds and cutoff values (listed in Table \ref{table:tracking_params}) from the values of our past work to better match the tracking to our data set and to improve the subjective quality of the tracking.
When considering ``contiguous'' or ``neighboring'' pixels, the present work
uses only the four above/below/left/right pixels and does not include the four
diagonal connections (which requires a large increase in the value of
$n_\text{exp}$), to avoid cases in which an identified \bp\ consists of
two separate regions with only a very tenuous, diagonal connection between them..
Our most significant change over our past work is the new expansion criterion of step 2.
This change significantly improves detection of the edges of each \bp, supporting our current analysis of shapes and sizes.
In comparing \bp\ boundaries drawn with our previous and current approaches, we
find from manual inspection that in the majority of cases the two criteria
produce comparable boundaries, with each approach at times performing better in
the subjective judgement of the viewer.
However, we also find a minority of cases in which our new approach produces significantly more plausible boundaries, whereas we did not identify any cases in which the new approach performed significantly worse.

\subsection{The Identified Boundaries}
\label{sec:new-boundaries}

\begin{figure*}[p]
	\centering
	\includegraphics[trim=0 .45cm 0 0,clip]{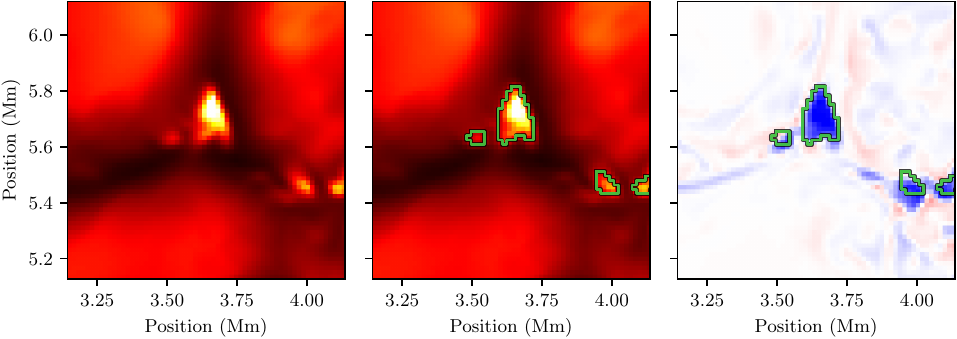}
	\includegraphics[trim=0 .34cm 0 0,clip]{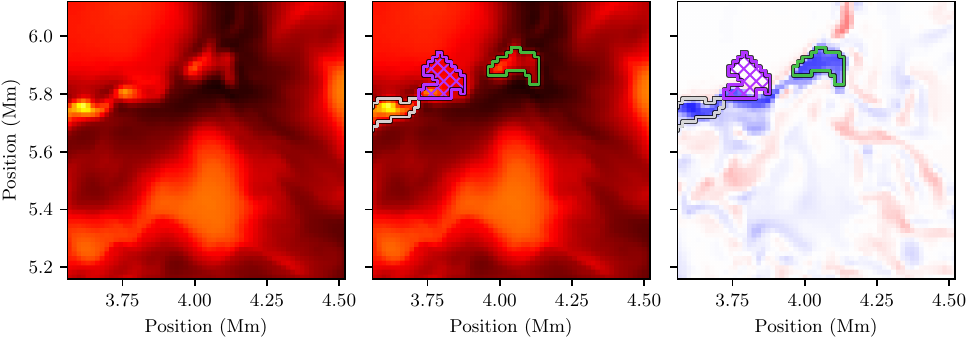}
	\includegraphics{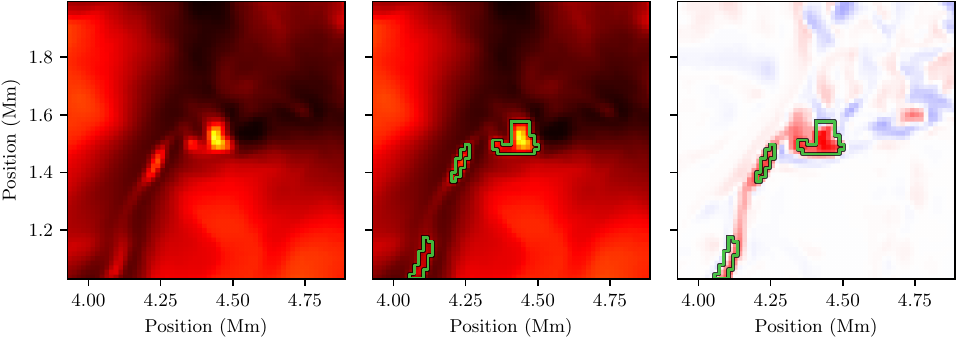}
	\caption{Sample of identified \bp s (continued on the next page). These \bp s
	were selected from a sample of 30 randomly-chosen \bp s to illustrate
	different shapes and scenarios. The left column provides an unobstructed view
	of the intensity pattern, and the center and right columns show \bp\
	boundaries overlaid on intensity and vertical magnetic field strength at
	$\tau=1$, respectively. Green lines mark the boundaries of accepted features.
	Magenta lines (with crosshatches) mark regions rejected by our false-positive
	rejection step. White lines mark regions rejected due to the minimum-lifetime
	constraint. Features rejected for other reasons are not present in these
	samples. Color maps range from 0.7~(black) to 1.75~(white) times the mean
	intensity for the intensity maps, and -2~(red) to 2 (blue)~kG for magnetic
	field strength. An animated version of the first row is available in
	the online journal. The 38-second animation shows the evolution of multiple \bp s
	over 382~s of simulation time, at a cadence of 2~s per frame.}
	\label{fig:tracking-bp-sample}
\end{figure*}

\begin{figure*}[p]
	\figurenum{\ref{fig:tracking-bp-sample}}
	\centering
	\includegraphics[trim=0 .34cm 0 0,clip]{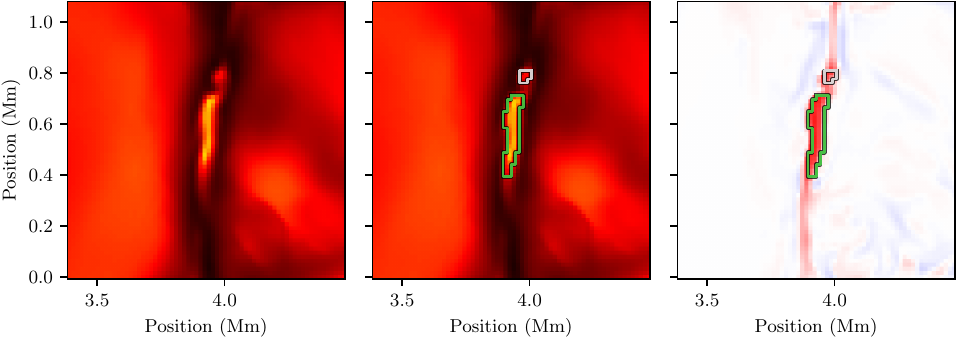}
	\includegraphics[trim=0 .45cm 0 0,clip]{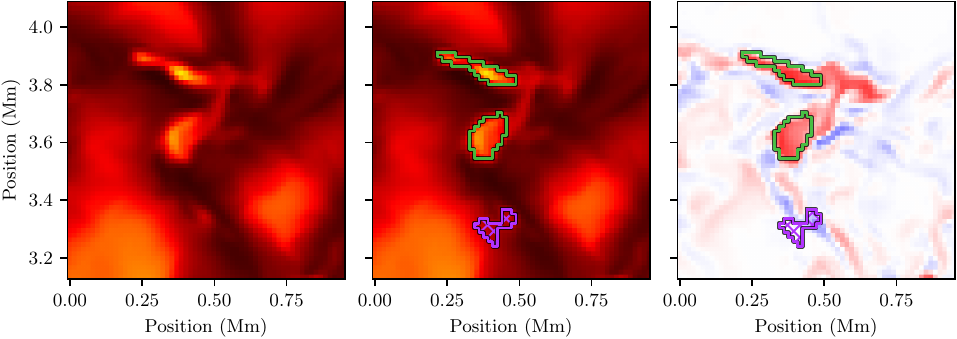}
	\includegraphics[trim=0 .45cm 0 0,clip]{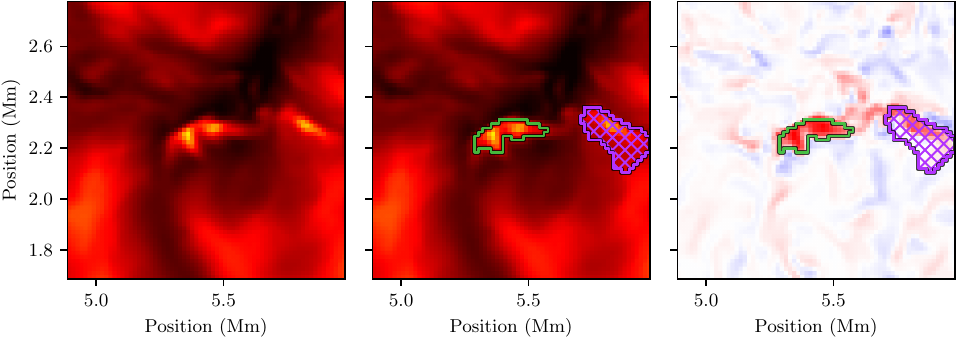}
	\includegraphics{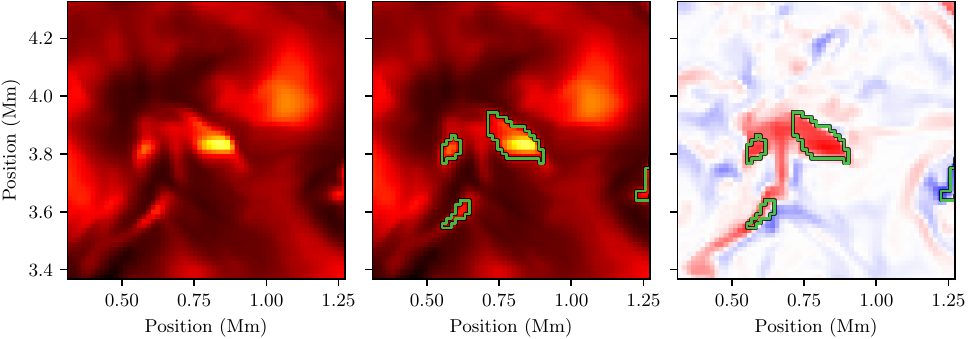}
	\caption{Sample of identified \bp s (cont.)}
\end{figure*}

In Figure~\ref{fig:tracking-bp-sample} we show a sample of identified \bhp\ boundaries.
Across these samples it can be seen that the boundaries typically align quite
well with the boundary that one would draw by eye.
In the included animation---which shows a variety of conditions, including splits, mergers, tracking imperfections, and a variety of both accepted and rejected features---it can be seen that the identified outlines very closely follow the evolving \bhp\ shape and tend to be very stable from frame to frame.
Additionally, comparison with the maps of vertical magnetic field strength shows that \bhp\ boundaries are typically, but not always, well-aligned with the boundaries of the magnetic enhancement, though it can also be seen that the flux concentration sometimes extends in a weakened form to other pixels not identified as part of the \bp.
It can also be seen that weak-field regions not associated with a white-light enhancement are abundant.
(These relations between intensity and magnetic flux at the smallest scales are relations that should be verified observationally with DKIST, and not necessarily accepted as fact based on these simulations. It is also important to note that the intensity maps represent a line-of-sight integration, while the field-strength maps show only the value along the $\tau=1$ surface, and so in any case differences between the two maps are expected.)

\subsubsection{\BP\ Statistics}
\label{sec:new-boundary-statistics}

\begin{figure*}[t]
	\centering
	\includegraphics[width=0.99\textwidth]{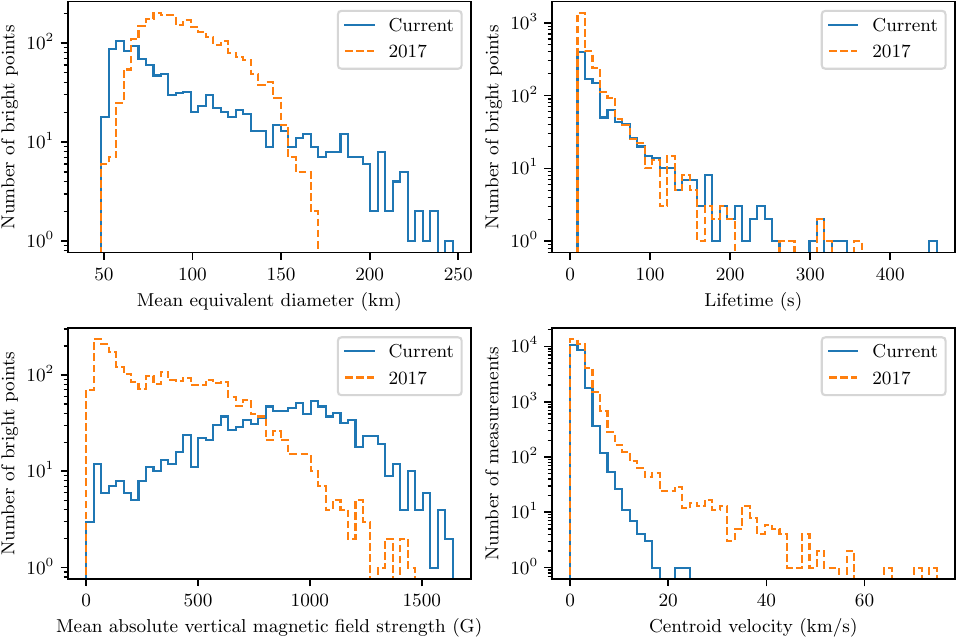}
	\caption{Distributions of \bp\ properties using our 2017 tracking and our current tracking. Equivalent diameter is the diameter of a circle of the same area as an identified \bp, and we plot the mean equivalent diameter across the lifetime of each \bp. Mean absolute vertical magnetic field strength is the mean of the absolute value of each pixel across the lifetime of the \bp\ and is measured at the $\tau=1$ surface. Centroid velocities are measured and plotted in between each pair of subsequent time steps.}
	\label{fig:tracking-distributions}
\end{figure*}

\begin{figure}[!ht]
	\centering
	\includegraphics[width=\columnwidth]{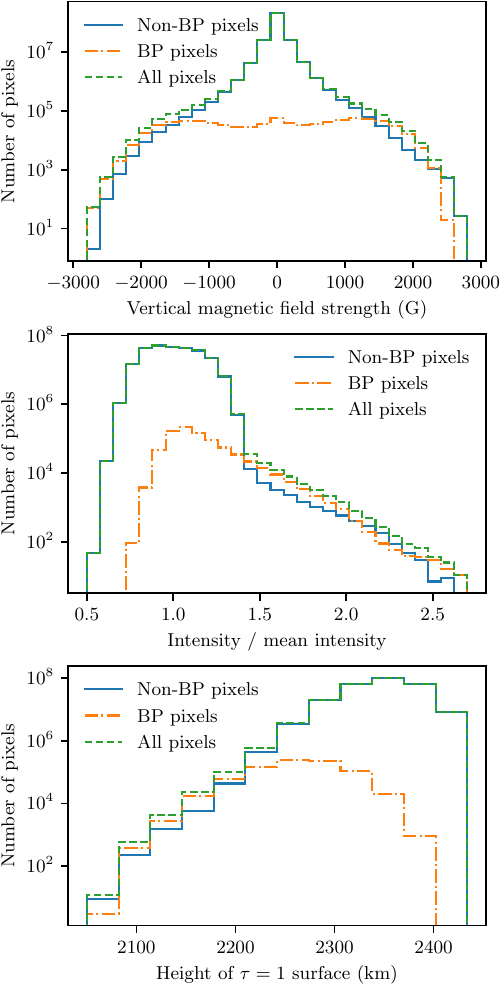}
	\caption{Distributions of pixel values for \bp s and full images. For each quantity, we show the distributions of the values of each pixel included in a \bp\ across all \bp s and all snapshots (``BP pixels''), the values of each pixel not included in a \bp, and the values of all pixels. Vertical magnetic field strength is measured at the $\tau=1$ surface. The height of the $\tau=1$ surface, computed independently at each pixel for a vertical line of sight, is measured relative to the bottom of the simulation box (which has a total height of $\sim 4$~Mm).}
	\label{fig:tracking-pixel-distributions}
\end{figure}

We identify 1,064 \bp s in the \muram\ data cube.
In Figure~\ref{fig:tracking-distributions} we show the distribution of \bhp\ sizes, lifetimes, and centroid velocities.
We include the distributions using our current tracking approach and, for
comparison, distributions produced by applying the tracking algorithm of our past
work \citep{VanKooten2017}.
It can be seen that larger \bp s are produced by our new approach (driven directly by the increase in $A\max$, $d\max$, and $n_\text{exp}$, though the effect of increasing $n_\text{exp}$ is partially to compensate for disallowing diagonal pixel connectivity).
Lifetimes with the new tracking are very slightly skewed toward longer lifetimes, suggesting that the updated feature identification makes the inter-frame linking more robust.
Centroid velocities are significantly reduced by the new tracking, which we interpret as being caused by the increased frame-to-frame consistency of the identified \bhp\ boundaries and therefore a reduction in random centroid ``jitter''---this was a strong motivation for the contour-based expansion step.
The distribution of per-\bhp\ mean vertical magnetic field strengths is shifted toward higher values with our updated tracking.
This may be because our new approach is more selective in the features it
identifies, and the more marginal features that are now rejected may be less
likely to be true or strong-field \bp s.
Mean values of these measured parameters with the current tracking are
95~km for equivalent diameter, 42~s for lifetime, 855~G for absolute vertical
magnetic field strength, and 1.8~km~s$^{-1}$ for centroid velocity.

It can be seen in Figure~\ref{fig:tracking-distributions} that the diameter
distribution peaks at the smallest allowed size and tapers off with a
rather consistent slope (in log space) to $\sim$140~km, after which the
slope flattens until reaching the largest sizes in the detected population.
This broken slope suggests our population of \bp s may be described well by a
two-component model (indeed, we produce a plausible fit with a two-Gaussian
model), consistent with the findings of \citet{BerriosSaavedra2022}, who
present observational evidence for the possibility of a two-population nature
for \bp s.

Figure~\ref{fig:tracking-pixel-distributions} shows the distribution of per-pixel values within our identified \bp s.
It can be seen that these values skew heavily toward strong vertical magnetic field strengths, high intensities, and low depths of continuum formation.
(Recall that \bp s appear bright because the high magnetic pressure offsets gas pressure and reduces gas density, lowering the $\tau=1$ surface to deeper, hotter plasma.)
Additionally, the clear majority of strong-field pixels are included in identified \bp s (though not the majority of flux: only 3.8\% of unsigned vertical magnetic flux is within a \bp).
Of note, however, these distributions do still include significant numbers of pixels with values that would not be expected for the ``ideal'' \bp, namely weak fields, low intensity, and high continuum formation.
While imperfect tracking no doubt contributes some of these values, others are due to the fact that, while the intensity enhancement and the
magnetic flux concentration are related and coupled phenomena, there is no
exact relation between the two quantities.
With this caveat in mind, the strong-field nature of most detected \bhp\ pixels
suggests that the intensity enhancements do serve as a useful proxy of the
fine structure of flux concentrations, even at these small scales.
(Observational validation of this claim at the highest resolutions must
await DKIST analysis, whereas this has been well demonstrated at medium
scales, e.g. \citealp{Ishikawa2007}.)

In \citet{VanKooten2021a} we explore in greater depth the properties of the identified \bp s and the impacts our algorithmic changes have on our results.
We show that our \bhp\ boundaries, on an individual basis, are at locations where the optical depth, intensity, and vertical magnetic field strength vary in ways characteristic of the edge of a flux tube.
We show that we get largely similar results from applying our tracking code (with few modifications) to photospheric slices of vertical magnetic field strength to track the magnetic flux enhancements that correspond to flux tubes.
We also show that our algorithmic changes significantly reduce the amount of centroid ``jitter,'' resulting in power spectra of centroid motion that are much more consistent in slope across a range of frequencies.

\section{Measuring Wave Modes}
\label{chap:wave-modes}

\subsection{Overview}
\label{sec:overview}

In this section, we present our technique for connecting thin-tube wave modes to \bhp\ observations.
In this method, we ``unroll'' the outlines of \bp s in polar coordinates, which we then fit with a sum of sinusoids, each representing a shape perturbation of a specific order $n$.
Treating changes in these fitted outlines as describing changes in the cross-sectional shape of a flux tube, we connect them to a range of MHD thin-tube wave modes, and we estimate an upward energy flux associated with each.
We assume that changes in the \bhp\ boundary are strictly caused by, and
directly connectable to, motions of the flux-tube wall, and that these changes
occur at one single height along the
strictly-vertical flux tube.
These assumptions are not guaranteed to hold (and in fact, it would be astonishing if they were universally true), but they are reasonable extensions of the assumptions in traditional centroid tracking.

We focus solely on the edges of the \bp, which can be measured in broadband observations at the highest possible resolution and cadence.
The edges are relatively sharp and easy to track, and we take them as a proxy for the edge of the overlying flux tube.
We disregard the intensity pattern within the \bp, which is more difficult to interpret physically.
It is very difficult in such an analysis, if not impossible, to fully incorporate the complex dynamics of \bp s and the mechanisms by which the visible-light enhancement connected to plasma flows, and we do not claim to do so in the present work.
Instead, we emphasize that our approach is a proof-of-concept, to be developed
further in future work and with the benefit of spectropolarimetric DKIST
observations (which will allow an understanding of the degree of
correspondence between flux-tubes and \bhp\ shapes and dynamics at the smallest scales).

We are not the first to consider incorporating the shapes of \bp s to more accurately treat their behavior and evolution (see, e.g., \citealp{Keys2020}, who proposed modeling \bp s as ellipses), but we believe that our approach of connecting the shapes of resolved \bp s to estimated flux-tube wave energy fluxes provides useful new insights.

\subsection{Method}
\label{sec:derivation}

\subsubsection{Summary of Linear Oscillation Properties}

The properties of small first-order wave-like perturbations have been studied extensively, and we will use much of the notation of \citet{Spruit1982} and \citet{Edwin1983}.
We assume a zeroth-order (unperturbed) flux tube that is cylindrical, with radius $r_0$, and oriented vertically in the solar photosphere.
We will use cylindrical coordinates $(r, \phi, z)$ to describe positional variations inside ($r < r_0$) and outside ($r > r_0$) the tube.
We assume the background magnetic field is always pointed along the $z$-axis of the cylinder.
The pressure, mass density, and field strength are $P_0$, $\rho_0$, and $B_0$ inside the tube, and $P_e$, $\rho_e$, and $B_e$ outside.
We also define the interior and exterior (adiabatic) sound speeds and \Al\ speeds in the standard way,
\begin{equation}
  c_0 \, = \, \sqrt{\frac{\gamma P_0}{\rho_0}}
  \qquad
  \mbox{and}
  \qquad
  c_e \, = \, \sqrt{\frac{\gamma P_e}{\rho_e}},
\end{equation}
\begin{equation}
  V_{{\rm A}0} \, = \, \frac{B_0}{\sqrt{4\pi\rho_0}}
  \qquad
  \mbox{and}
  \qquad
  V_{{\rm A}e} \, = \, \frac{B_e}{\sqrt{4\pi\rho_e}},
\end{equation}
and we further assume that the exterior region is field-free, such that $B_e \approx V_{{\rm A}e} \approx 0$.

The standard derivation for compressive MHD waves begins by assuming that the divergence of the velocity fluctuations behaves as
\begin{equation}
  \nabla \cdot \vec{v} \, = \, \Delta \, = \,
  R(r) \, \exp \left( i \omega t \, + \, i n \phi \, + \, ikz \right),
  \label{eq:Delta}
\end{equation}
where $R(r)$ is a yet-undetermined function, and the azimuthal mode number $n$ is the primary identifier of the mode of the oscillation.
We will solve a dispersion relation for the frequency $\omega$ as a function of the axial wavenumber $k$.
Two other key quantities derived by \citet{Spruit1982} and \citet{Edwin1983} are the radial wavenumbers for the interior and exterior of the tube:
\begin{equation}
  m_0^2 \, = \,
  \frac{(k^2 c_0^2 - \omega^2)(k^2 V_{{\rm A}0}^2 - \omega^2)}
  {(k^2 c_{{\rm T}0}^2 - \omega^2)(c_0^2 + V_{{\rm A}0}^2)}
\end{equation}
and
\begin{equation}
  m_e^2 \, = \,
  \frac{(k^2 c_e^2 - \omega^2)(k^2 V_{{\rm A}e}^2 - \omega^2)}
  {(k^2 c_{{\rm T}e}^2 - \omega^2)(c_e^2 + V_{{\rm A}e}^2)},
\end{equation}
where the tube speed $c_{\rm T}$ is defined inside and outside the tube as
\begin{equation}
  c_{{\rm T}0} \, = \, \frac{c_0 V_{{\rm A}0}}
  {\sqrt{c_0^2 + V_{{\rm A}0}^2}}
  \quad \mbox{and} \quad
  c_{{\rm T}e} \, = \, \frac{c_e V_{{\rm A}e}}
  {\sqrt{c_e^2 + V_{{\rm A}e}^2}}
  \, .
\end{equation}
Note that $m_e^2$ is assumed to always be positive, while $m_0^2$ can be positive or negative. When $m_0^2 < 0$, we define $\mu_0^2 = -m_0^2 > 0$.

The solutions for $R(r)$ take the form of Bessel functions.
Both the sign of $m_0^2$ and the boundary conditions determine which type of Bessel functions to use:
\begin{equation}
  R(r) \, = \, A \left\{
  \begin{array}{ll}
     I_n (m_0 r) \, , &
     r < r_0 \,\,\, \mbox{and} \,\,\, m_0^2 > 0 \\
     J_n (\mu_0 r) \, , &
     r < r_0 \,\,\, \mbox{and} \,\,\, \mu_0^2 > 0 \\
     K_n (m_e r) \, , &
     r > r_0 
  \end{array}
  \right.
  \label{eq:Rbess}
\end{equation}
where $A$ is a normalization constant with units of s$^{-1}$.
\citet{Spruit1982} derived the expressions for all three components of the first-order velocity perturbation.
Leaving off the exponential term in Equation~\eqref{eq:Delta}, these are
\begin{equation}
  v_z \, = \, -A \, \frac{c^2}{\omega^2} \, ik {\cal B}_n \, ,
  \label{eq:vdef_z}
\end{equation}
\begin{equation}
  v_r \, = \, A \, \frac{\omega^2 - k^2 c^2}{m^2 \omega^2}
  \frac{d}{dr} {\cal B}_n \, ,
  \label{eq:vdef_r}
\end{equation}
and
\begin{equation}
  v_{\phi} \, = \, iA \, \frac{\omega^2 - k^2 c^2}{m^2 \omega^2} \,
  \frac{n}{r} {\cal B}_n
  \, ,
  \label{eq:vdef_phi}
\end{equation}
where the general Bessel function ${\cal B}_n$ must be replaced with a function and argument as in Equation~\eqref{eq:Rbess} above.
Note also that the interior or exterior versions of $m$ and $c$ must be substituted as well.

The last key quantity is the dispersion relation for $\omega(k)$.
It is given by ensuring that both $v_r$ and the total pressure are continuous across the boundary at $r_0$.
\citet{Edwin1983} specifies two versions for two distinct cases.
For ``surface waves'' ($m_0^2 > 0$),
\begin{equation}
\begin{split}
  \rho_0 (k^2 V_{{\rm A}0}^2 - \omega^2) \, m_e \,
  \frac{K'_n (m_e r_0)}{K_n (m_e r_0)} \\
  = \,\,
  \rho_e (k^2 V_{{\rm A}e}^2 - \omega^2) \, m_0 \,
  \frac{I'_n (m_0 r_0)}{I_n (m_0 r_0)}
\end{split}
\end{equation}
and for ``body waves'' ($m_0^2 < 0$),
\begin{equation}
\begin{split}
  \rho_0 (k^2 V_{{\rm A}0}^2 - \omega^2) \, m_e \,
  \frac{K'_n (m_e r_0)}{K_n (m_e r_0)} \\
  = \,\,
  \rho_e (k^2 V_{{\rm A}e}^2 - \omega^2) \, \mu_0 \,
  \frac{J'_n (\mu_0 r_0)}{J_n (\mu_0 r_0)}
\end{split}
\end{equation}
where ${\cal B}'_n (m r_0)$ is $(d/dx) {\cal B}_n (x)$ evaluated at $x = m r_0$.

\subsubsection{The Thin Tube Approximation}

We will assume that the arguments of the Bessel functions are all $\ll 1$ (the thin-tube limit), and we will evaluate the appropriateness of this assumption shortly.

In this limit, the surface-wave and body-wave dispersion relations produce the same expressions for $\omega^2/k^2$, and these expressions depend on the value of $n$.
For $n=0$ (``sausage-mode'' waves),
\begin{equation}
  V_{\rm ph}^2 \, = \, \frac{\omega^2}{k^2} \, = \, c_{{\rm T}0}^2
  \, .
\end{equation}
For $n \geq 1$ (general ``kink-mode'' waves),
\begin{equation}
  V_{\rm ph}^2 \, = \, \frac{\omega^2}{k^2} \, = \, c_{\rm k}^2 \, ,
\end{equation}
where
\begin{equation}
  c_{\rm k}^2 \, = \,
  \frac{\rho_0 V_{{\rm A}0}^2 + \rho_e V_{{\rm A}e}^2}
       {\rho_0 + \rho_e}
\end{equation}
defines the kink speed $c_{\rm k}$.
The phase velocity is constant, indicating these are dispersionless waves with $V_{\rm gr} = V_{\rm ph}$.

With these phase velocities in hand, we now calculate values for many of these quantities and evaluate the appropriateness of the thin-tube approximation.
Drawing from the model of \citet{Cranmer2005}, at the photosphere of a typical intergranular \bp\ surrounded by a field-free granule, we have $B_0 = 1430$~G, $\rho_e = 3 \times 10^{-7}$~g~cm$^{-3}$, and a density contrast of $\rho_0 / \rho_e = 0.316$.
If the temperature is the same inside and outside the tube, then
\begin{align*}
  V_{{\rm A}0} &= \mbox{13.1 km~s$^{-1}$} \, , \\
  V_{{\rm A}e} &= 0 \, , \\
  c_0 = c_e &= \mbox{8.13 km~s$^{-1}$} \, ,
\end{align*}
where we also assume $T = 5770$~K and the mean atomic weight $\mu = 1.2$ for the mostly neutral photosphere.
The gas pressures ($P = \rho k T / \mu m_{\rm H}$) inside and outside
the tube are roughly $3.765 \times 10^4$ and $1.190 \times 10^5$
dyne~cm$^{-2}$, respectively.
Thus,
\begin{align*}
  c_{{\rm T}0} &= \mbox{6.91 km~s$^{-1}$} \, , \\
  c_{{\rm T}e} &= 0 \, , \\
  c_{\rm k} &= \mbox{6.42 km~s$^{-1}$} \, .
\end{align*}
The field-free limit for the external medium simplifies the evaluation of $m_e$, since in this case
\begin{equation}
  m_e^2 \, \approx \, k^2 \left( 1 - \frac{V_{\rm ph}^2}{c_e^2} \right)
\end{equation}
which is indeed always positive for the above values.
For $n=0$, $m_e \approx 0.528 k$, and for $n \geq 1$, $m_e \approx 0.614 k$.

The interior radial wavenumber $m_0$ can be written as
\begin{equation}
  m_0^2 \, = \, k^2 \,
  \frac{(c_0^2 - V_{\rm ph}^2)(V_{{\rm A}0}^2 - V_{\rm ph}^2)}
  {(c_{{\rm T}0}^2 - V_{\rm ph}^2)(c_0^2 + V_{{\rm A}0}^2)} \, ,
  \label{eq:m0vph}
\end{equation}
and we examine two cases:
\begin{itemize}
\item
For $n=0$, with $V_\text{ph} = c_\text{T0}$, the denominator of Equation~\eqref{eq:m0vph} is zero.
However, \citet{Edwin1983} show numerical solutions where $c_{{\rm T}0}^2 < V_{\rm ph}^2 < c_0^2$ (with $V_{\rm ph}$ very close to $c_\text{T0}$ for small values of the argument to the Bessel functions) producing $m_0^2 < 0$ and requiring $J_n$ Bessel functions.
Later, we will show that $m_0$ (or rather, $\mu_0$) is not present in the final equations for the $n=0$ modes, so that the resulting very large magnitude for $m_0$ is not physically relevant.
\item
For $n \geq 1$, $V_{\rm ph} = c_{\rm k}$, and Equation~\eqref{eq:m0vph} produces $m_0 \approx 1.448 k$.
Since $m_0^2 > 0$, the $I_n$ Bessel functions are used for kink modes.
\end{itemize}
To evaluate the suitability of the thin-tube limit, we take $r_0 = 50$~km (for a flux tube of order 100~km across) and $V_{\rm ph} = 6.5$~km~s$^{-1}$ for the $n\ge1$ modes.
Taking $m \approx k$, the Bessel-function argument $x \approx k r_0 = \omega r_0 / V_{\rm ph}$, which depends on the frequency.
Representative timescales for \bhp\ evolution range from about 2 to 20 minutes.
This produces a span from $x \approx 0.04$ (for the longer period) to $x \approx 0.4$ (for the shorter period).
For the $n=0$ mode, taking $V_{\rm ph} = 1.1 c_{\rm T0}$, those same representative timescales produce values from $x \approx 0.02$ (for the longer period) to $x \approx 0.2$ (for the shorter period).
We see that the thin-tube approximation is not suitable for any faster flux-tube motions, but it is certainly appropriate for representative motions over the bulk of a \bp 's life.

\subsubsection{Kink Modes: Velocity Amplitudes and Energy Fluxes}
\label{sec:n1p-fluxes}

Here we analyze the $n \geq 1$ (kink) waves, with $n=0$ waves deferred for later.
First, we derive the velocity field both inside and outside the tube.
In the thin-tube limit, $I'_n (x) \approx n I_n (x) / x$, so that Equations \eqref{eq:vdef_r} and \eqref{eq:vdef_phi} give $v_\phi = i v_r$.
Additionally, the axial (i.e., vertical) velocity amplitude is negligible for a kink-mode wave, since
\begin{equation}
  \frac{v_z}{v_r} \, = \, -i N_0^2
  \left( \frac{c_0^2}{V_{\rm ph}^2 - c_0^2} \right)
  \frac{kr}{n} \, ,
\end{equation}
where $N_0 = m_0 / k$ is a dimensionless number.
For the example numbers given above, $N_0 \approx 1.448$ and $|v_z / v_r| \approx 5.57 kr/n$.
We are treating $n$ values of order one, and the thin-tube limit sets $kr \ll 1$.
Thus, we can ignore the contributions of $v_z$ motions to the total energy fluxes of modes with $n \geq 1$.

We aim to derive the full eigenfunctions for $v_r(r)$ and $v_{\phi}(r)$ for different values of $n$.
The constants $A$ given in Equations \eqref{eq:vdef_z} through \eqref{eq:vdef_phi} are not continuous across the boundary of the tube, but $v_r$ itself must be continuous.
Thus, we define $A_0$ inside the tube and $A_e$ outside the tube, with
\begin{equation}
  v_r (r < r_0) \, = \, \frac{A_0}{m_0}
  \left( 1 - \frac{c_0^2}{V_{\rm ph}^2} \right)
  \frac{n \, (m_0 r)^{n-1}}{2^n \, n!}
\end{equation}
\begin{equation}
  v_r (r > r_0) \, = \, -\frac{A_e}{m_e}
  \left( 1 - \frac{c_e^2}{V_{\rm ph}^2} \right)
  \frac{n \, 2^{n-1} \, (n-1)!}{(m_e r)^{n+1}}
\end{equation}
where, outside the tube, we use $K'_n (x) \approx -n K_n (x) / x$.
Equating the two velocities at $r_0$, and assuming for now that $c_0 = c_e$, we find
\begin{equation}
  \frac{A_e}{A_0} \,\, = \,\, -\frac{m_e}{m_0} \,
  \frac{(m_0 r_0)^{n-1} (m_e r_0)^{n+1}}{2^{2n-1} \,
  (n-1)! \, n!} \,\, .
\end{equation}
However, if we choose to normalize everything by the value of $v_r$ at the tube boundary (which we label $V_0$), the expression simplifies to
\begin{equation}
  v_r(r) \,\, = \,\, V_0 \left\{
  \begin{array}{ll}
    (r/r_0)^{n-1} \, , & \mbox{if $r < r_0$} \\
    (r_0/r)^{n+1} \, , & \mbox{if $r > r_0$} \, .
  \end{array}  \right.
  \label{eq:vrsimple}
\end{equation}
Note that the complete expression for $v_r(r,\phi,z,t)$ is the above function multiplied by $\exp ( i \omega t + i n \phi + ikz )$.
It can be seen that the $n=1$ mode displays uniform radial motion throughout the body of the flux tube (consistent with its nature of offsetting the entire tube cross section).
Higher-$n$ modes are increasingly concentrated at the tube boundary.

As an aside, we showed earlier that inside the flux tube, $v_{\phi} = i v_r$.
However, outside the flux tube, $v_{\phi} = -i v_r$.
Thus, if $v_r$ is forced to be continuous at the boundary, then $v_{\phi}$ must be discontinuous (by changing sign).
This was discussed by \citet{Goossens2009}.

The instantaneous kinetic energy density carried by a wave with complex amplitudes is defined as
\begin{equation}
  \varepsilon \, = \, \frac{1}{2} \rho \langle
    \textrm{Re}(\vec{v}) \cdot \textrm{Re}(\vec{v}) \rangle \, ,
  \label{eqn:epsilon}
\end{equation}
where $\rho$ is the zero-order density ($\rho_0$ inside the tube and $\rho_e$ outside the tube), $\textrm{Re}({\bf v})$ is the real part of vector {\bf v}, and angle brackets denote an average over one wave period.
For sinusoidal time dependence, this is equivalent to
\begin{equation}
  \varepsilon \, = \, \frac{1}{4} \rho
  ( \vec{v} \cdot \vec{v}^{\ast} )
\end{equation}
\citep[see, e.g.,][]{Mihalas1984,Walker2005}.
Note that the above expressions for $\varepsilon$ are fully spatially dependent, but since we have used Equation~\eqref{eq:Delta}, the only remaining spatial variation is in the $r$ direction.
Thus, if
\begin{equation}
  v_r (r,\phi,z,t) \,\, = \,\, V_0 \, f(r) \,
  \exp \left( i \omega t \, + \, i n \phi \, + \, ikz \right) \, ,
  \label{eqn:vr-form}
\end{equation}
where we define $f(r)$ as the dimensionless quantity to the right of the braces in Equation~\eqref{eq:vrsimple}, then
\begin{equation}
  {\bf v} \cdot {\bf v}^{\ast}
  \,\, = \,\,
  v_r v_r^{\ast} \, + \, v_{\phi} v_{\phi}^{\ast} 
  \,\, = \,\,
  2 v_r v_r^{\ast}
  \,\, = \,\,
  2 | V_0 \, f(r) |^2 \,\, .
  \label{eq:v-dot-v}
\end{equation}

Traditionally, one multiplies $\varepsilon$ by the group velocity to obtain a kinetic energy flux (with units W~m$^{-2}$) of
\begin{equation}
  F \,\, = \,\, \frac{1}{2} \rho | V_0 \, f(r) |^2 \, V_{\rm gr} \,\, .
\end{equation}
However, because each isolated flux tube contains both internal and external fluctuations, we integrate over the horizontal plane to obtain the power (in W) associated with that one tube.
Later, we can sum this power over all detected \bp s and divide by the total area under study to produce a spatially-averaged energy flux across a patch of the solar surface.
We write the power (associated with the kinetic energy) as
\begin{equation}
  \dot{E}_{\rm K} \,\, = \,\, \int dA \,\, F
  \,\, = \,\, 2\pi \int_0^{\infty} dr \,\, r \,\, F
  \,\, .
\end{equation}
Note that the two relevant integrals over $f^2$ are identical:
\begin{equation}
\begin{aligned}
  &\int_0^{r_0} dr \,\, r \, \left( \frac{r}{r_0} \right)^{2n-2} \\
  = \,\,
  &\int_{r_0}^{\infty} dr \,\, r \, \left( \frac{r_0}{r} \right)^{2n+2}
  \,\, = \,\,
  \frac{r_0^2}{2n} \, ,
\end{aligned}
\end{equation}
so the power is given by
\begin{equation}
  \dot{E}_{\rm K} \,\, = \,\, \frac{\pi r_0^2}{2n} \,
  \left( \rho_0 + \rho_e \right) \, V_0^2 \, V_{\rm gr}
  \,\, .
  \label{eqn:Edot}
\end{equation}
Usually, for linear MHD waves, the kinetic energy is exactly half of the total energy carried by all forms of variability (with the other half being magnetic and thermal energy; see Section~2.5 of \citealp{Kulsrud2005}), so $\dot{E}_{\rm tot} = 2 \dot{E}_{\rm K}$.
In any case, if we can measure $V_0$ for each mode $n \geq 1$, we can compute each mode's contribution to the total power
\citep[see also][]{Goossens2013,VanDoorsselaere2014}.

\subsubsection{Sausage Modes: Velocity Amplitudes and Energy Fluxes}
\label{sec:sausage-modes}

We now consider the separate case of $n=0$ modes.
The velocity amplitudes are given by Equations \eqref{eq:vdef_z} through \eqref{eq:vdef_phi} with appropriate choices of the Bessel functions.
Note that $v_{\phi} = 0$ for these circularly-symmetric oscillations.
In the thin-tube limit, the vertical velocity amplitudes are
\begin{equation}
\begin{aligned}
  v_z (r < r_0) \,\, &= \,\,
  -A_0 \, \frac{c_0^2}{\omega^2} \, ik \, J_0 (\mu_0 r) \\
  &\approx \,\,
  -i A_0 \, \frac{c_0^2}{c_{{\rm T}0}^2} \, \frac{1}{k}
  \label{eq:sausageVZIN}
\end{aligned}
\end{equation}
\begin{equation}
\begin{aligned}
  v_z (r > r_0) \,\, &= \,\,
  -A_e \, \frac{c_e^2}{\omega^2} \, ik \, K_0 (m_e r) \\
  &\approx \,\,
  i A_e \, \frac{c_e^2}{c_{{\rm T}0}^2} \, \frac{1}{k}
  \, \ln (m_e r) \, ,
  \label{eq:sausageVZOUT}
\end{aligned}
\end{equation}
and the radial velocity amplitudes are
\begin{equation}
\begin{aligned}
  v_r (r < r_0) \,\, &= \,\,
  A_0 \, \left( \frac{\omega^2 - k^2 c_0^2}{-\mu_0^2 \, \omega^2} \right) \, 
  \frac{d}{dr} J_0 (\mu_0 r) \\
  &\approx \,\,
  A_0 \, \left( \frac{c_{{\rm T}0}^2 - c_0^2}{c_{{\rm T}0}^2} \right)
  \, \frac{r}{2}
  \label{eq:sausageVRIN}
\end{aligned}
\end{equation}
\begin{equation}
\begin{aligned}
  v_r (r > r_0) \,\, &= \,\,
  A_e \, \left( \frac{\omega^2 - k^2 c_e^2}{m_e^2 \, \omega^2} \right) \, 
  \frac{d}{dr} K_0 (m_e r) \\
  &\approx \,\,
  -A_e \, \left( \frac{c_{{\rm T}0}^2 - c_e^2}{c_{{\rm T}0}^2} \right)
  \, \frac{1}{m_e^2 \, r} \, .
\end{aligned}
\end{equation}
Note also that because
\begin{equation}
  \frac{d}{dr} J_0 (\mu_0 r) \,\, \approx \,\,
  -\frac{\mu_0^2 \, r}{2}
\end{equation}
the explicit dependence of $v_r$ on $\mu_0$ (in the $r < r_0$ case) cancels out in the thin-tube limit, so its value need not be computed.

As with the $n \ge 1$ case, we can write the radial velocity in terms of its value at the tube boundary:
\begin{equation}
  v_r(r) \,\, = \,\, V_0 \left\{
  \begin{array}{ll}
    r/r_0 \, , & \mbox{if $r < r_0$} \\
    r_0/r \, , & \mbox{if $r > r_0$}
  \end{array}  \right.
\end{equation}
By requiring $v_r$ be continuous at $r=r_0$, we can find expressions for $A_0$ and $A_e$ (assuming $c_0 = c_e$ as before):
\begin{align}
  A_0 \, &= \, \frac{2 V_0}{r_0} \left(
  \frac{c_{{\rm T}0}^2}{c_{{\rm T}0}^2 - c_0^2} \right) \, , \\
  A_e \, &= \, - m_e^2 \, r_0 V_0  \left(
  \frac{c_{{\rm T}0}^2}{c_{{\rm T}0}^2 - c_0^2} \right) \, .
\end{align}
Note that in the thin-tube limit, the dominant velocity is the vertical $v_z$ inside the flux tube, and this quantity is also spatially constant inside the tube.

To compute the upward kinetic energy flux, we assume that $\vec{v} \cdot \vec{v}^{\ast} \approx |v_z^2|$ in the interior of the tube, and $\vec{v} \cdot \vec{v}^{\ast} \approx 0$ outside, so that
\begin{equation}
  F \,\, \approx \,\, \left\{
  \begin{array}{ll}
    \frac{1}{4} \rho_0 \, | v_z^2 |  \, V_{\rm gr} \, ,
      & \mbox{if $r < r_0$} \\
    0 \, , & \mbox{if $r > r_0$}
  \end{array}  \right.
\end{equation}
and
\begin{equation}
  \dot{E}_{\rm K} 
  \,\, = \,\, 2\pi \int_0^{\infty} dr \, r \, F
  \,\, = \,\,
  \frac{\pi r_0^2}{4} \, \rho_0 \, | v_z^2 |  \, V_{\rm gr} \, ,
  \label{eq:EKn0}
\end{equation}
which agrees with Equation~(24) of \citet{Moreels2015}.
(Recall that $\dot{E}_{\rm K}$ must be doubled to account for all forms of energy.)
We discuss next how we infer $v_z$ from observations.

\subsubsection{Connecting Shape Changes to Waves}
\label{sec:shape-changes-to-waves}

To connect measured shape changes to this traditional framework of thin-tube waves, we take our \bhp\ shapes and ``unroll'' their outlines into polar coordinates relative to their centroid, producing a function $r(\phi)$ that describes the outline.
We then fit $r(\phi)$ with a sum of sinusoids
\begin{equation}
	\begin{split}
	R(\phi) &= r_0 + \sum_{n=2}^{n\max} \left[ A_n \cos\left(n\phi\right) - B_n \sin\left(n \phi \right) \right] \\
	&= r_0 + \sum_{n=2}^{n\max} \textrm{Re}\left[\left(A_n + iB_n\right) e^{i n \phi}\right]\,,
	\end{split}
	\label{eqn:sum-of-sinusoids}
\end{equation}
where Re indicates the real part of the expression.
This represents an unperturbed circle of radius $r_0$ with radial perturbations for wave modes $n=2$ through $n\max$, a chosen parameter.
(The $n=1$ mode, representing centroid motion, is treated separately, as $r(\phi)$ is defined relative to the centroid location.)
$A_n$ and $B_n$ represent the fitted parameters (together representing an amplitude and phase).

After fitting each observed shape across a \bp 's lifetime, we have time series of $r_0(t)$, $A_n(t)$ and $B_n(t)$.
Differentiating these time series (and thereby differentiating Equation~\eqref{eqn:sum-of-sinusoids}), we obtain measured values
\begin{equation}
\begin{split}
	v_r(\phi) &= \dot{r}_0 - \sum_{n=2}^{n\max} \left[ \dot{A}_n \sin\left(n\phi\right) + \dot{B}_n \cos\left(n \phi \right) \right] \\
	v_r(\phi) &= \dot{r}_0 - \sum_{n=2}^{n\max} \textrm{Re}\left[\left(\dot{B}_n + i\dot{A}_n\right) e^{in\phi}\right] \,.
	\label{eqn:v-sum-of-sinusoids}
\end{split}
\end{equation}
Each term in the summation represents a perturbing wave of the form of Equation~\eqref{eqn:vr-form}, at some fixed height $z$ (introducing an arbitrary phase offset) and with $r$ fixed at the \bhp\ radius $r_0$ and therefore $f(r) = f(r_0) = 1$.
The expression $\dot{B}_n + i\dot{A}_n$ therefore fills the role of $V_0$, and can be carried through to Equation~\eqref{eqn:Edot}.
(As shown earlier, $v_z$ is negligible for these modes, and $v_\phi = i v_r$ is accounted for in the progression to Equation~\eqref{eqn:Edot}.)
The previous derivations presented $V_0$ as a constant, while here it is a time series of values due to the multi-frequency nature of real oscillations.
An RMS of that $V_0$ series can provide a mean effective amplitude and energy flux, while a power spectrum can probe the frequency content of the observations.

Attempting this fitting exposes an immediate problem, that many \bhp\ shapes do not unroll into single-valued functions $r(\phi)$ (i.e., the \bhp\ edge crosses a $\phi$ value multiple times).
To address this, we replace these problematic shapes with approximations that are single-valued functions.
We term our chosen solution ``shrinkwrapping,'' in which we draw a circle of points, uniformly-spaced in $\phi$ and at a large radial distance centered on the \bhp\ centroid, we draw each point radially in until it first touches the actual \bhp\ edge, and we treat these points as our approximating shape.
Some \bp s are problematic in that their centroid lies outside their boundary, and so a suitable stopping condition must be defined as each point is drawn radially in.
For points that reach the centroid before touching the \bhp\ border, we stop the points at $r=0$.

Our approach measures perturbations to the assumed equilibrium, circular shape of the \bp\ and treats the measured centroid of the \bp\ in any one image as the true, unperturbed centroid.
Since $n=1$ perturbations (in the small-amplitude limit) represent offsets of the instantaneous centroid relative to the unperturbed centroid, they do not fit in this framework, and so we measure them in the traditional way (see, e.g., \citealp{VanKooten2017}) of measuring the velocity of the measured centroid location between successive frames, which serves as $V_0$ for this mode.
(The $n=1$ mode is the only mode that perturbs the centroid location.)

The $n=0$ mode also requires a different approach, as this mode has its dominant velocity in $v_z$, while only $v_r$ can be directly inferred from changes in the shape of the flux tube.
At the tube boundary (where $|v_r| = V_0$), we can use Equations (\ref{eq:sausageVZIN}) and (\ref{eq:sausageVRIN}) to write
\begin{equation}
  \frac{v_z}{v_r}
  \,\, = \,\,
  - i \frac{2}{k r_0} 
  \left( \frac{c_0^2}{c_{{\rm T}0}^2 - c_0^2} \right)
  \,\, = \,\,
  i \frac{2}{k r_0} 
  \left( \frac{c_0^2 + V_{{\rm A}0}^2}{c_0^2} \right) \, .
  \label{eq:defouw}
\end{equation}
The above expression reproduces Equation~(12) of \citet{Defouw1976} and confirms that in the thin-tube limit ($k r_0 \ll 1$), $v_z > v_r$.

We write the time series of measured \bhp\ areas as
\begin{equation}
	A(t) \, = \, \pi r_0^2 \, + \, 2 \pi r_0 \frac{V_0}{\omega} \, \cos(\omega t + \psi) \, .
	\label{eqn:M00}
\end{equation}
The constant term $\pi r_0^2$ is the mean value of the area $A(t)$.
We take the RMS after subtracting the mean (equivalent to taking the standard deviation, which we denote as $\sd(\;\cdot\;)$),
\begin{equation}
	\sd(A(t)) \, = \, \sqrt{2} \, \pi r_0 \frac{V_0}{\omega}
	\label{eqn:sd-M00}
\end{equation}
(using the fact that the standard deviation of a sinusoid is $1/\sqrt{2}$ times its amplitude).
With this expression, an effective value of $V_0$ can be computed from a time series of area measurements.
$V_0$ is the magnitude of the radial velocity oscillations at the flux-tube boundary, but we require a value for the dominant vertical velocities.
With Equation~\ref{eq:defouw}, we find
\begin{equation}
\begin{split}
	\left| v_z \right|
	\, &= \, \frac{\sqrt{2} \; \sd(A(t)) }{\pi r_0^2} \, \frac{\omega}{k} \left( \frac{c_0^2 + V_{A0}^2}{c_0^2} \right) \\
	&= \, \frac{\sqrt{2} \; \sd(A(t)) }{\pi r_0^2} \left( \frac{V_{A0}^2}{c_{\rm T0}} \right),
	\label{eqn:vmag-n0}
\end{split}
\end{equation}
where we have used $\omega / k \, = \, c_{\rm T0}$, a constant.
This expression serves as an estimate of the amplitude of $v_z$.
An observed sequence of $A(t)$ and $v_r$ values will, in fact, be a sum of these sinusoidal oscillations at varying frequencies.
Taking advantage of this sinusoidal nature, $|v_z| / \sqrt{2}$ can be taken to be an effective RMS of $v_z$, and it can be used directly in Equation~\eqref{eq:EKn0} to compute a mean upward energy transfer rate over time for the sum of the individual wave components.

If we desire a power spectrum of the energy flux, however, a different approach must be taken.
Each observed $v_r$ value will be due to the contributions of those multiple sinusoids of varying frequencies, and it derives not from the amplitudes of those sinusoids but from individual values taken from any point along those sinusoidal curves.
Additionally, because of the presence of multiple frequency components, the frequency-dependent Equation~\eqref{eq:defouw} cannot be used to convert these instantaneous $v_r$ values into $v_z$ values.
(It is this requirement to convert the measurable $v_r$ values into the dominant $v_z$ values that sets the $n=0$ mode apart from the $n\ge1$ modes.)
Instead, we re-write Equation~\eqref{eqn:M00} as
\begin{equation}
	A(t) \, = \, \pi r_0^2 \, + \, 2 \pi r_0 r_1(t) \, ,
\end{equation}
which expresses the area of the flux-tube cross section as the area of the unperturbed cross section and an area perturbation due to a small perturbation $r_1(t)$ in radius.
Taking the time derivative yields
\begin{equation}
	\frac{d}{dt} \, A(t) \, = \, 2 \pi r_0 \frac{d}{dt} \, r_1(t) \, = \, 2 \pi r_0 v_r(t)  \, ,
	\label{eqn:vr_time_series}
\end{equation}
where we have used $v_r(r=r_0) = dr_1(t)/dt$, which follows from the radial position of the tube edge, $r(t) = r_0 + r_1(t)$.
Equation~\eqref{eqn:vr_time_series} directly connects the derivative of a time series of $A(t)$ measurements to a time series of $v_r$ measurements (assuming a constant value of $r_0$, which we take to be the average of $r(t) = \sqrt{A(t)/\pi}$).
We then take the Fourier transform of the $v_r$ time series, which then allows us to re-scale each frequency bin from $v_r$ to $v_z$ using Equation~\eqref{eq:defouw}, with $k=\omega/c_{\rm T0}$.
We then take an inverse-Fourier transform to recover a time series of estimated $v_z$ values.
Repeating for each \bp\ then allows us to produce an average $v_z$ (or $\dot{E}$) power spectrum for the $n=0$ mode.

\subsubsection{Temporal smoothing}
\label{sec:outlines-smoothing}

One potential hazard in this approach is that of over-discretization of the \bp 's shape changes by a very rapid cadence.
Consider, for example, the case of uniform rotation of an ellipsoidal \bp.
While the underlying motion is smooth, pixels must enter or leave the identified \bp\ at discrete points in time, and this causes a degree of ``jumpiness'' in the identified outlines of the \bp.
To address this, we apply a temporal smoothing to our outline fitting with a window size of 5 frames.
For each time step in a \bp 's lifetime, we fit Equation~\eqref{eqn:v-sum-of-sinusoids} to a set of $r(\phi)$ points containing the outline at that time step, with a weight of 1 for each point, and the outlines identified in the two frames before and two frames after, with weights of $2/3$ for the nearer frames and $1/3$ for the further frames.
This ensures that pixels' influence gradually enters and leaves the feature, that the fitted perturbations vary smoothly with time, and that high-frequency noise in the fits is suppressed.
Our choice of this smoothing window is discussed further in Section~\ref{sec:effect-of-smoothing}.

\section{Results}
\label{sec:results}

We present here results from applying this technique to \bp s identified in simulated observations from the \muram\ simulation described in Section~\ref{sec:2s-muram}.
An archive of our analysis code and data is available \citep{VanKooten2023a}.

\begin{figure*}[p]
	\centering
	\includegraphics{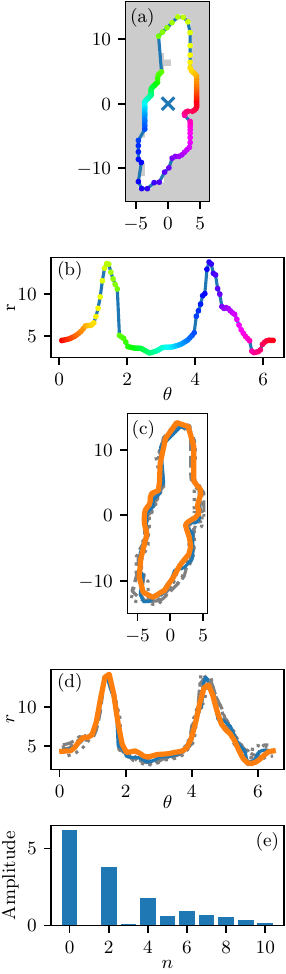}
	\includegraphics{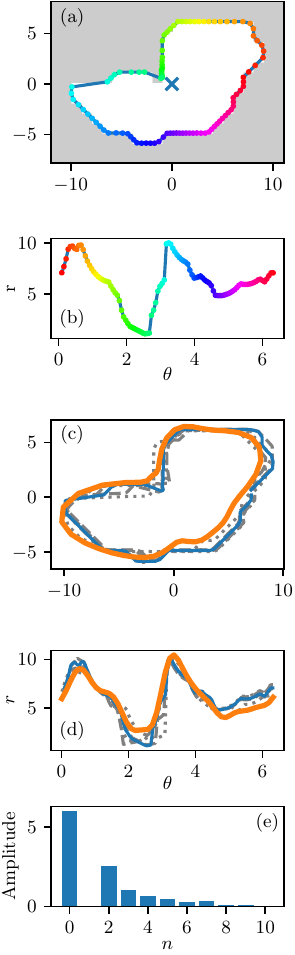}
	\includegraphics{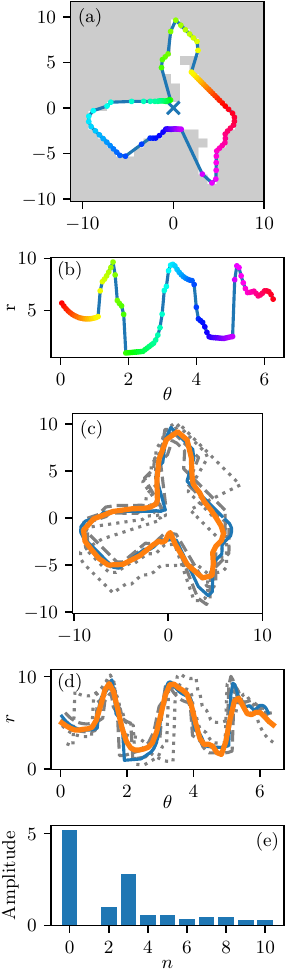}
	\caption{A sample of three identified \bp s (each of the three columns), illustrating the fitting process. In each column, (a)~shows the identified \bp\ pixels in white, the shrinkwrapped outline as rainbow dots, and the centroid as an ``x''; (b)~shows the unrolled outline in polar coordinates; (c)~shows the fitted outline in orange and the source outlines being fit, with solid blue lines indicating the outline with weight 1, gray dashes indicating the neighboring frames with weights $2/3$, and gray dots indicating the more distant neighboring frames with weights $1/3$; (d)~shows the fitted and source outlines in polar coordinates; and (e)~shows the amplitudes of each fitted component ($\left|A_n + iB_n\right|$, using the terminology of Equation~\eqref{eqn:sum-of-sinusoids}). The $n=1$ mode is not represented on these plots, as it must be defined relative to an unperturbed centroid location, and so it is not a single-frame quantity. The third column shows a particularly interesting case, in which the $n=3$ component is dominant, rather than the $n=2$ component that dominates most \bp s. This example also illustrates well how some shape fidelity can be lost to shrinkwrapping.}
	\label{fig:bp-example-fits}
\end{figure*}

\subsection{Fitted Outlines}

In Figure~\ref{fig:bp-example-fits} we show a sample of \bhp\ shapes fit using the method of \ref{sec:derivation}, including an illustration of the shrinkwrapping process and the amplitudes of each fitted term.
It can be seen that these three samples, which were chosen somewhat arbitrarily, are reasonably-well approximated by the fitted shapes.
While we show fitted amplitudes for one point in time for three \bp s, it is the time derivative of these amplitudes that determines the wave energy flux.

\begin{figure*}[t]
	\centering
	\includegraphics{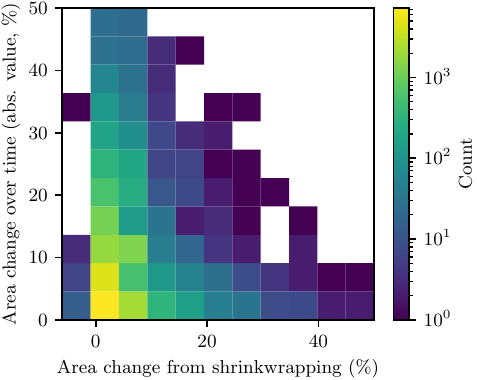}
	\includegraphics{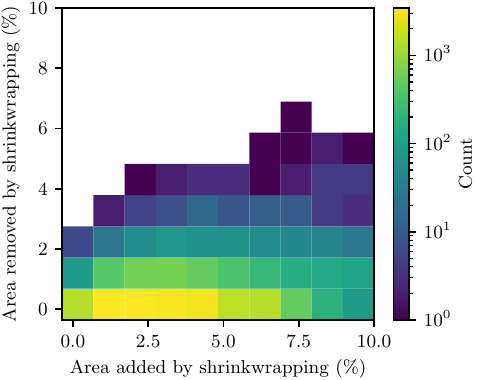}
	\caption{Impact of shrinkwrapping on \bp\ areas. On the left, we show a 2D histogram showing, for each \bp\ and each timestep, the change in that \bp 's area due to shrinkwrapping and the change in its (un-shrinkwrapped) area over the following timestep. It can be seen that in almost all cases, the area change due to shrinkwrapping is smaller than that which occurs naturally over time. On the right, we show the area of regions added to the \bp\ and of regions removed from the \bp\ by the shrinkwrapping process.}
	\label{fig:shrinkwrapping-effect}
\end{figure*}

To understand the impact of the shrinkwrapping process, which is necessary to produce single-valued $r(\phi)$ fits, we show this procedure's impact in Figure~\ref{fig:shrinkwrapping-effect}.
We compute the area contained within our identified \bhp\ edges before and after shrinkwrapping, and find that this area change is typically very small (a few percent), and it also follows a more narrow distribution than the distribution of area changes between two consecutive frames, indicating that the impact of shrinkwrapping is smaller than the actual \bhp\ shape changes that we seek to measure.
One could imagine a situation in which shrinkwrapping both adds area to the \bp\ on one side and subtracts area on another side in roughly equal amounts, conspiring to produce significant shape changes but little total area change.
We thus also separately measure the total amount of area added and subtracted from each \bp\ due to shrinkwrapping.
We do this by overlaying the two shape outlines, both before and after wrapping, on a very fine pixel grid and summing the pixels that are contained in one but not the other of the two shapes.
We find that area subtractions tend to be very small or non-existent, with area changes almost exclusively being area additions.
This is consistent with the general behavior of shrinkwrapping being to remove concave portions of the \bhp\ shape.

\begin{figure*}[t]
	\centering
	\includegraphics{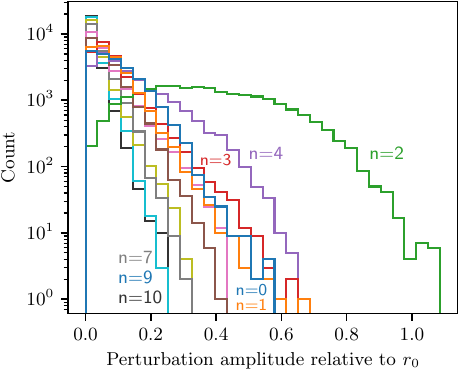}
	\includegraphics{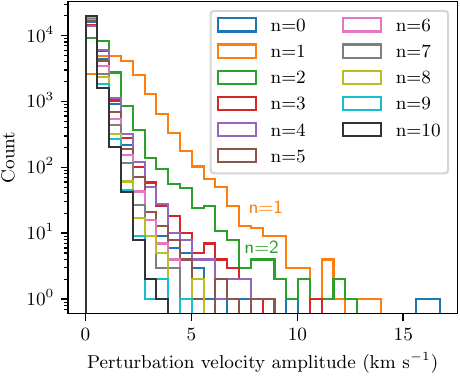}
	\caption{Distribution of fitted perturbation amplitudes across all \bp s and time steps. On the left are the per-mode radial perturbation amplitudes ($\left|A_n + iB_n\right|$ in the notation of Equation~\eqref{eqn:v-sum-of-sinusoids}) expressed relative to $r_0$. The $n=1$ amplitude is measured as the offset in centroid position between two time steps. On the right are the radial velocity amplitudes (the time derivatives of the radial perturbations).}
	\label{fig:fitted-outline-amplitudes}
\end{figure*}

In Figure~\ref{fig:fitted-outline-amplitudes}, we show the distribution of the amplitudes of our fitted perturbations.
\Bp s are confined within the long, thin lanes between granules, which tends to produce \bp s with long, thin shapes.
This is clearly reflected in the distribution of $n=2$ perturbation amplitudes (and, to a lesser extent, $n=4$), which displays a strong offset from zero as well as much larger typical values compared to the other modes.
When taking the time derivative of these perturbations to find the radial velocity of the \bhp\ edge (which is the quantity we connect to energy flux), the dominance of the $n=2$ mode is diminished and we see the largest values in the $n=1$ mode.
This leads directly to the dominance of that mode in energy flux, which we show next, which is amplified by a $1/n$ scaling between velocity amplitudes and energy fluxes.

\subsection{Energy Fluxes}
\label{sec:energy-fluxes}

\begin{table*}[t]
	\centering
	\begin{tabular}{l|ccccccccccc}
		Quantity & $n=0$ & $n=1$ & $n=2$ & $n=3$ & $n=4$ & $n=5$ & $n=6$ & $n=7$ & $n=8$ & $n=9$ & $n=10$ \\
		\hline
		\rule{0pt}{.9\normalbaselineskip}
		Mean flux ($10^3$ W~m$^{-2}$) & 6.9 & 26 & 4.5 & 1.6 & 1.1 & 0.57 & 0.4 & 0.27 & 0.21 & 0.14 & 0.11 \\
		\rule{0pt}{0cm}
		Mean $\dot{E}_\textrm{tot}$ ($10^{16}$ W) & 2.2 & 8 & 1.4 & 0.5 & 0.34 & 0.18 & 0.13 & 0.084 & 0.067 & 0.045 & 0.036 \\
	\end{tabular}
	\caption{Mean vertical flux and $\dot{E}_\textrm{tot}$ values for each wave mode. The fluxes are averaged over the full simulation domain, and the mean $\dot{E}_\textrm{tot}$ values are across all \bp s and time steps. The large difference in the magnitudes of these quantities is due to the very small filling factor of \bp s. The fluxes are corrected as described in Section~\ref{sec:energy-fluxes}.}
	\label{table:flux-values}
\end{table*}

\begin{figure}[t]
	\centering
	\includegraphics{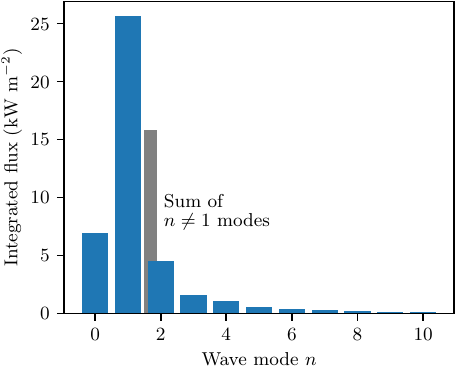}
	\caption{Per-mode energy fluxes integrated over the entire spatial and temporal domain of the simulation (that is, the ``mean flux'' values of Table \ref{table:flux-values}). The gray bar indicates the total flux of all modes $n\ne1$, for comparison to the long-studied $n=1$ flux.}
	\label{fig:integrated-fluxes}
\end{figure}

We described methods for inferring energy fluxes in Section~\ref{sec:n1p-fluxes} for $n\ge1$ modes and Section~\ref{sec:sausage-modes} for $n=0$ modes.
For each \bp, we use these techniques to produce time series of $\dot{E}_\textrm{tot}$ values, indicating the rate of vertical energy transfer through the flux tube, integrated over the area of the \bp\ (and surrounding areas, for the $n\ge1$ modes where external perturbations are significant).
These values can be integrated across the lifetime of each \bp, and then summed across all \bp s to produce a total amount of energy transferred across the simulation run-time.
We then divide by the simulation duration and area to produce a spatially- and temporally-averaged vertical energy flux across the simulation domain.
The mean flux values we compute, as well as mean values for $\dot{E}_\textrm{tot}$, are shown in Table \ref{table:flux-values} for each wave modes, and the fluxes are also represented in Figure~\ref{fig:integrated-fluxes}.
These fluxes have all been scaled by a correction factor as described at the end of this sub-section.

The $n=1$ mode is clearly dominant in both measures, as was evident in Figure~\ref{fig:fitted-outline-amplitudes} where the $n=1$ mode was dominant in perturbation velocities.
But the $n=1$ mode is only dominant by a factor of $4$ compared to the next-strongest mode ($n=0$), and its flux is only about 50\% larger than the sum of the $n\ne1$ fluxes.
This indicates that $n\ne1$ modes (primarily the $n=0,2$ modes) may make a significant contribution to the energy flux budget.
We believe these wave modes are therefore a very compelling target for future study, as they have the potential to significantly increase the potency of wave--turbulence models of coronal heating.
It is important to note, though, that the upward propagation of these modes through the rapidly-varying plasma properties of the chromosphere and transition region must be modeled before any implications to the coronal energy budget become clear.

In computing these fluxes, we have first applied a filtering step to remove \bp s that may be problematic.
We remove any \bp s whose fitted outline $r(\phi)$ is negative at any $\phi$, which is approximately 1\% of \bp s.
These negative values typically indicate a severe violation of the assumption of small perturbations on a cylindrical flux tube.
We remove about 8\% of \bp s for which, at any point in the identified \bhp\ lifespan, the shrinkwrapping process changes the total area by more that 15\% (i.e. cases where the post-shrinkwrapping shape is not a good approximation for the true shape).
We finally remove the 22\% of \bp s which show an area change of more than 35\% at any point in their lifetime.
These tend to be small \bp s, where an otherwise-small area change represents a large percentage difference.
Nonetheless, such a large relative change may cause us to infer a large, impulsive wave flux that is more likely a result of the limited resolution than a true effect.
28\% of all \bp s fall into these categories.
Since these removed \bp s are identified \bp s, just with shape histories that are more difficult to treat with this approach, ignoring them can cause an under-estimation of the total wave-energy flux.
To account for this, we have divided our reported fluxes by the fraction of \bp s that survive these filters---effectively assuming that each ignored \bp\ has an energy flux equal to the average of the accepted \bp s.
As these filters primarily remove \bp s with extremely large, impulsive changes in area, the removed \bp s tend to have correspondingly-large inferred energy fluxes, and so this imputation may represent a conservative under-estimate of the energy flux for any incorrectly-rejected \bp s.

\subsection{Flux Comparisons}
\label{sec:flux-comparisons}

These flux values can be put in context by comparing them to the upward Poynting flux in the \muram\ simulation.
We calculated this flux, averaged horizontally and across 19 data cubes evenly spaced through the simulation.
We found that this flux is negative immediately below the average $\tau=1$ surface, rises rapidly to a peak value of 33~kW~m$^{-2}$ at a height of 200~km above the photosphere, and drops to a value of zero at the simulation's upper boundary.
We interpret the negative fluxes below the photosphere as being associated with the strong convective downflows in the lanes where \bp s exist, as well as downward-propagating waves excited near the photosphere.
We interpret the drop-off toward the top of the simulation as damping by the upper boundary condition.
We therefore take the peak value of 33~kW~m$^{-2}$ as the most proper comparison point, where the influence of both convective downflows and boundary-condition damping is minimized.
When comparing the Poynting flux to the wave energy fluxes above, it is important to remember that the wave fluxes were arrived at by doubling the kinetic energy flux, since MHD-wave energy transport is half kinetic energy and half ``other'' or potential energy---the latter half includes magnetic and thermal energy \citep[see][]{Kulsrud2005}.
Appropriate values for the magnetic energy flux that can be compared with the Poynting flux are therefore at most half of the wave fluxes above (e.g. 3.5, 13, and 2.3~kW~m$^{-2}$ for the $n=0,1,2$ modes, respectively).
These values fit together within the Poynting flux budget (which is not expected to consist solely of flux tube waves above \bp s), indicating both that the order of magnitude of our fluxes is reasonable, and that the Poynting flux does not show our fluxes to be over-estimates.

In our past work \citep{VanKooten2017}, we computed an $n=1$ flux value from the observational \bhp\ centroid velocities of \citet{Chitta2012}.
Disregarding for now the factor used in that work to account for expected wave reflection, and using the $\rho$ and $B$ values of Section~\ref{chap:wave-modes}, these observational velocities produce a comparison flux of 10.7~kW~m$^{-2}$ for the $n=1$ mode.
Our value is approximately 2.5 times larger, an effect that may be due to increased spatial resolution allowing finer motions to be resolved and more \bp s to be identified (see Section~\ref{sec:blurring-data}).

It is worth comparing these estimated energy fluxes to those required to maintain coronal temperatures.
As a simple approach to estimating how all of these wave modes propagate to the corona, we can multiply each estimated flux by $(1-\mathcal{R})/(1+\mathcal{R})$, where $\mathcal{R}$ is a reflection coefficient.
\citet{Cranmer2005} modeled the propagation of the $n=1$ mode from the photosphere to the corona, and found typical values of $\mathcal{R}$ of $\sim 0.9$ below the transition region (with dramatically-reduced reflection above the transition region).
Simply applying this factor to all wave modes as a zeroth-order estimation produces energy fluxes at the base of the corona of 360, 1,400, and 240~W~m$^{-2}$ for the $n=0,1,2$ modes, respectively.
These fluxes compare favorably to the classical fluxes required to maintain coronal temperatures of \citet{Withbroe1977}, i.e., 300 and 800~W~m$^{-2}$ for quiet-Sun and coronal hole regions, emphasizing that these wave modes, as part of a wave-turbulence-based heating model, are plausible as a source of coronal heating.

\subsection{Spectra}
\label{sec:spectra}

\begin{figure}[t]
	\centering
	\includegraphics{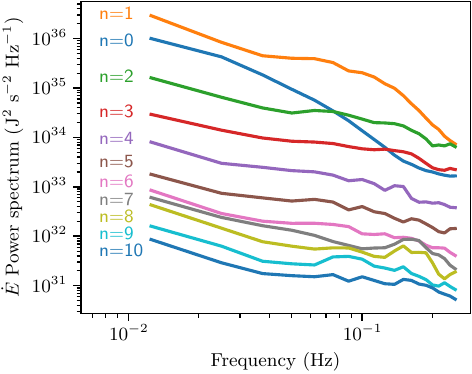}
	\caption{Power spectra of the $\dot{E}$ time series for each wave mode.}
	\label{fig:modes-power-spectrum}
\end{figure}

In Figure~\ref{fig:modes-power-spectrum} we plot averaged power spectra of the $\dot{E}$ time series (choosing to show spectra for $\dot{E}$ rather than velocities to account for the mode-dependent relation between the two quantities).
We restrict this analysis to only the 58 \bp s which we tracked for at least 40 frames, to ensure adequate frequency content in each sequence as we produce our average spectrum.
We employ Welch's method \citep{Welch1967}, dividing these time series into a total of 283 overlapping, Hann-windowed, 40-frame segments (using zero padding at the ends) and we average the spectra of each of these segments.
The $n\ge2$ spectra generally have very similar and featureless spectra, while the $n=0,1$ spectra show steepening at higher frequencies.
In general, all the spectra show broad frequency content, with no one clearly-dominant frequency range.

\subsection{Effect of Smoothing}
\label{sec:effect-of-smoothing}

\begin{figure}[t]
	\centering
	\includegraphics{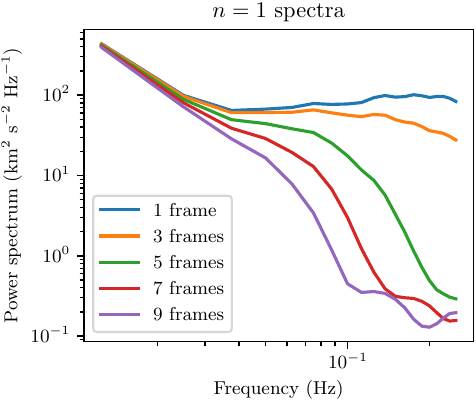}\\
	\vspace{.1in}
	\includegraphics{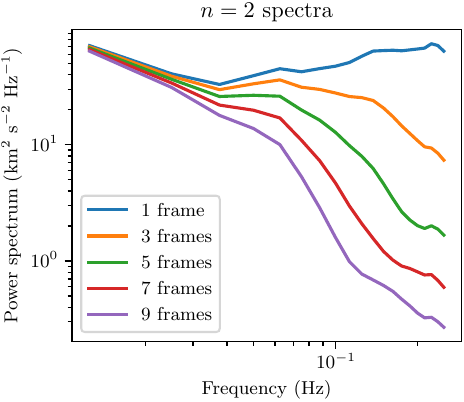}
	\caption{Power spectra of $n=1$ and 2 waves for different temporal smoothing windows. The temporal smoothing window width is given for each spectrum, with weights dropping linearly from 1 at the window center to 0 at the first snapshot outside the window. One time step is 2~s.}
	\label{fig:smoothing-spectra}
\end{figure}

We consider here the effect of the temporal smoothing we employ in this work, implemented via the pixel weights described in Section~\ref{sec:outlines-smoothing}.
In Figure~\ref{fig:smoothing-spectra} we show power spectra for the $n=1$ and 2 modes with various sizes of the temporal smoothing window.
The un-smoothed curve for both modes changes to an upward trend at the lowest frequencies, indicative of contamination from high-frequency ``jitter'' in the \bhp\ shape outlines (i.e. jitter due to the impulsive way pixels are added to or removed from identified features, as well occasional pixels that are added to and then rapidly removed (or vice versa) from an identified feature, if those pixel values remain close to a relevant feature-identification threshold).
This jitter was a concern discussed in the Appendix of \citet{VanKooten2017}.
A smoothing window of 3~time steps (6~s) can be seen to make a large reduction in this effect, while 5~time steps (10~s) is the smallest window that tends to ensure monotonicity.
This window size does tend to produce a significant drop in power at the highest frequencies, which may indicate excessive damping.
At and above this 5~time-step window size, spectral power begins to decrease in the middle of the spectrum, suggesting that any larger window may over-smooth the \bp s at all frequencies.
We therefore select a temporal smoothing window of 5~time steps, which seems to be a middle choice that removes high-frequency jitter while erring only modestly toward over-smoothing.

\subsection{Effect of Resolution and Blurring the Data}
\label{sec:blurring-data}

A relevant question in this ``preparing for DKIST'' context is how our analyses change if the \muram\ images are degraded to a resolution more comparable to pre-DKIST observations.
We investigate this by convolving the \muram\ images with an Airy disk of radius 100~km (or 6.25 \muram\ pixels).
We run our \bhp\ tracking on these blurred images, finding through manual inspection of a sample of \bp s that no modifications to our algorithm or its parameters are required for satisfactory tracking of the blurred \bp s.
The number of tracked and accepted \bp s is reduced by a factor of 3.3, from 1,064 to 324 \bp s.
The \bp s that are not detected in the blurred data tend to be difficult or impossible to detect by eye in the blurred data.
The average \bhp\ lifetime increases by a factor of 1.2, from 42~s to 52~s, and the average \bhp\ area increases by a factor of 1.6, from 36~px to 57~px, suggesting that it is predominantly small and short-lived \bp s that are lost.
This interpretation is confirmed by the size ratios of individual \bp s.
The 324 tracked \bp s in the blurred data account for 8,261 individual, single-frame outlines.
Of these outlines, 7,246 can be unambiguously linked to a single outline in the unblurred tracked results---that is, the blurred feature overlaps exactly one unblurred feature, and that unblurred feature overlaps exactly one blurred feature.
Among these one-to-one matches, the geometric mean of the ratios of the size of a \bp\ as detected in the blurred and unblurred tracking is 1.03, indicating that blurring does not typically affect the identified size of a \bp.
The increase in average \bhp\ size can therefore be interpreted as a loss of small \bp s.
(Only a third of the 324 \bp s in the blurred data show a one-to-one match that is consistent across their entire lifetime, so we do not attempt a similar analysis for the changes in lifetimes.)

\begin{figure}[t]
	\centering
	\includegraphics[width=\columnwidth]{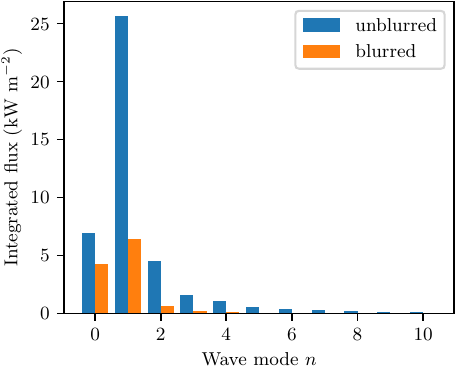}
	\caption{Energy fluxes computed from blurred and unblurred images.}
	\label{fig:blurring-fluxes}
\end{figure}

Following the methods described earlier, we compute energy fluxes using our blurred tracking results.
This reveals a mode-dependent reduction in flux, shown in Figure~\ref{fig:blurring-fluxes}.
The $n=0$ flux is reduced from 6.9 to 4.2~kW~m$^{-2}$, the $n=1$ flux from 26 to 6.4~kW~m$^{-2}$, and the $n=2$ flux from 4.5 to 0.67~kW~m$^{-2}$, with the higher modes all reduced to under 0.2~kW~m$^{-2}$.
This blurring produces an $n=1$ flux much more comparable to the value of 10.7 kW~m$^{-2}$ we computed from the observations of \citet{Chitta2012} (discussed in Section~\ref{sec:flux-comparisons}), suggesting that much of the increased $n=1$ flux we detect in the \muram\ simulations may be due to the presence of smaller \bp s that are only detectable in the higher-resolution data, and/or the ability to resolve finer motions.

The flux estimated from the blurred data is approximately 60\% of the unblurred $n=0$ flux, 25\% of the unblurred $n=1$ mode, 15\% of the unblurred $n=2$ mode, and 10\% for the higher modes, suggesting that the higher modes are most sensitive to shape information being lost at low resolution, which is to be expected.

\section{Discussion}
\label{sec:discussion}

In addition to applying these techniques to DKIST observations (which
we intend to do in future work), At least four factors must
be considered before moving forward with the estimated energy fluxes.

First is the ``true'' filling factor of \bp s.
Our tracking algorithm does not capture every \bp, and not every strong-field region corresponds to a visible \bp---this can be seen, for example, in Figure~\ref{fig:tracking-bp-sample}.
This means that we may be underestimating the total wave energy flux.
This can be resolved, in part, by further work refining the \bp\ tracking, but that still will not account for all vertical magnetic flux.
Our identified \bp s account for 42\% of pixels with $|B_z| > 1000$~G, 14\% of pixels with $|B_z| > 500$~G, and 4\% of all unsigned vertical flux.
(This distribution is shown in Figure~\ref{fig:tracking-pixel-distributions}.)
Thus, an ad-hoc correction factor of $1/0.42$ might be considered to approximately account for the energy flux in strong-field regions not included in the identified \bhp\ ensemble---regions which might be expected to experience wave driving very similar to the \bp s we measure.
An alternative way to approach this problem is to track magnetic flux concentrations directly in magnetograms, to apply our techniques to flux-tube cross sections inferred from those measurements, and to consider differences in the fluxes computed from \bhp\ tracking and flux-element tracking as a rough estimate of the methodological uncertainty present in each flux value, including, in part, the variability due to the difficulty in identifying all flux tubes.

Second, one must assess the degree to which flux tubes reach the corona, as opposed to bending over and re-entering the photosphere (e.g., through an opposite-polarity \bp).
In regions with mixed magnetic polarity, the latter scenario may be expected to occur with some frequency.
Flux tubes with a large vertical extent may therefore be more commonly rooted in large, unipolar network regions.
(The \muram\ simulation we use does not include the supergranular network, and
so this is an effect we are unable to test.)
An estimate of the fraction of \bp s reaching the corona might be obtained by using a potential field extrapolation from measurements of the average magnetic field within individual \bp s \citep[see, e.g.,][]{Close2003,VanBallegooijen2003,Wiegelmann2004}.

Third, the degree of propagation of these waves (accounting for both reflection and damping) must be understood before our estimated energy fluxes can be considered in a coronal context.
For $n=1$ flux-tube waves, \citet{Cranmer2005} modeled the wave propagation from the photosphere, through the heights where the flux tubes are expected to merge into a ``canopy,'' through the rapidly-changing plasma properties of the chromosphere and transition region, and into the corona and heliosphere.
They found that approximately 5\% of the wave flux reaches beyond the transition region.
(\citealp{Soler2019} find a similarly strong reduction in flux for the torsional \Al\ wave, a mode we do not consider in the present work.)
This factor cannot be directly transferred to the $n\ne1$ modes, but it is to
be expected that these other modes will also have a reduced fraction of their
flux succeed in reaching the corona.
An increase in the wave flux may also affect the rate at which waves
experience reflection or mode conversion at the sharp transition region
boundary \citep[see, e.g.,][]{Cranmer2023}.
Thus, for a variety of reasons, further modeling work must be undertaken before
the implications of our inferred $n\ne1$ wave-mode driving can be understood,
including the degree to which the energy budget for wave dissipation in the
corona is changed by the addition of these wave modes.
We hope that the present work provides new motivation for a thorough modeling
effort of these modes.

Fourth, this study should be repeated on high-resolution observations (which we intend to do with DKIST). While this work provides strong motivation for doing so, the present study only considers simulated images. While observations will of course provide stronger results, they will bring their own difficulties, such as accounting for viewing-angle and projection effects in the observations if far from disk-center, which may complicate both the identification of \bp s and the interpretation of their foreshortened shapes; and accounting for seeing effects and ensuring corrections do not significantly damp observed \bhp\ motions. Such factors will be addressed in future work.

In addition, future work might reassess our theoretical interpretation of the measured shape changes to account for the non-circular shape of \bp s.
Our approach assumed perturbations on a cylindrical flux-tube cross section, but many \bp s have more elliptical shapes due to their confinement in granular downflow lanes.
MHD waves in elliptical flux tubes have been considered in some scenarios
\citep{Aldhafeeri2021}, and such a treatment could increase the strength of our
analysis.
Further, not every \bhp\ shape is close to a circular (or elliptical)
shape, meaning the ``perturbations'' required to describe its shape may not be
low-amplitude. (See Figure~\ref{fig:fitted-outline-amplitudes} for the
distribution of amplitudes.) This
challenges our framework for interpreting those perturbations as waves, and
future work should consider whether large-amplitude perturbations should be
excluded from the analysis.

One interesting avenue for future exploration is to track flux-tube footpoints by tracking the corresponding magnetic flux enhancement as seen in photospheric magnetograms (i.e., to track the magnetic manifestation of the footpoint rather than the white-light manifestation).
This may be expected to more directly probe the cross-sectional shape of the magnetic flux tube.
However, magnetograms require spectropolarimetric observations that are much more demanding than the broadband observations that support \bhp\ measurements.
Tracking flux concentrations therefore requires a compromise in cadence, limiting the wave frequencies than can be probed, or in field-of-view, limiting the number of flux concentrations that can be seen.
The utility of flux-concentration tracking---even with DKIST---may therefore be limited to more specific case studies where the evolution of a few flux concentrations are compared to that of their corresponding \bp s, giving opportunities to constrain and validate the conclusions drawn from \bhp\ tracking.

\section{Conclusions}
\label{sec:conclusions}

In this paper, we have focused on preparing for DKIST observations of \bp s, which can resolve the shapes of \bp s at a level never before achieved.
These resolved shapes can provide new insights on wave--turbulence heating models; however, doing so requires analysis techniques beyond the centroid tracking that has traditionally been applied to unresolved \bp s to study $n=1$ waves.
We introduced what we believe to be a novel method for analyzing the shape changes of \bp s and connecting them to the modeled excitation of flux-tube waves of different modes.
We decompose the shapes of \bp s into a sum of sinusoids, and we connect variations in these sinusoids to wave-driven perturbations in the cross-section of a thin flux tube.
We considered $n=0$--10 modes, and our framework can be extended to higher-$n$ modes (though higher modes appear to be unimportant).

Our estimated energy fluxes, when applying this technique to simulated observations from a \muram\ simulation of DKIST-like resolution, show that the $n=1$ mode is dominant, but the $n=0$ and 2 modes make significant contributions to the wave energy budget, together being approximately half the $n=1$ flux.
This suggests that these wave modes may be significant contributors to wave heating of the coronal in quiet-sun regions, beyond the $n=1$ energy flux that has long been the focus of \bhp\ studies.

These estimated fluxes, though, are to be understood as existing at and immediately above the photosphere.
Further modeling must be done to understand if and how these wave modes propagate upward, where they dissipate, and thus whether they make significant contributions to the heating budget in the corona.
The results we present now provide a strong motivation for such modeling efforts, and also strongly motivate the application of these techniques to DKIST observations, which we intend to do soon.

DKIST's VBI instrument offers a diffraction limit matching the grid spacing of the \muram\ simulation we analyze presently and an even finer pixel scale, which will reduce the impact of the discretization of \bhp\ shapes (potentially reducing the necessity of our temporal smoothing step).
We thus anticipate that our technique will transfer well to observations and perhaps allow even cleaner inference of wave fluxes.

\begin{acknowledgements}

We thank Matthias Rempel for making available the results of his \muram\ runs, Piyush Agrawal for assisting in acquiring those data files, and colleagues (including S.V.'s dissertation committee) for fruitful discussions of this work at all stages.

S.V. was funded by NASA FINESST grant number 80NSSC20K1503 and National Science Foundation (NSF) grant number 1613207.

This research has made use of NASA’s Astrophysics Data System Bibliographic Services.

\end{acknowledgements}

\software{Matplotlib \citep{Hunter2007} version 3.7.1 \citep{matplotlib371}, NumPy \citep{Harris2020} version 1.23.5, SciPy \citep{Virtanen2020} version 1.10.1 \citep{scipy1101}}


\end{document}